\documentclass[aps,prf,preprint,showpacs,superscriptaddress,longbibliography]{revtex4-1}
\usepackage[utf8]{inputenc}
\usepackage[english]{babel}
\usepackage{amsmath,upgreek}
\usepackage{amsfonts}
\usepackage{amssymb}
\usepackage[left=2cm,right=2cm,top=2cm,bottom=2cm]{geometry}
\usepackage{graphicx}
\usepackage{color}
\usepackage{bm}

\begin{document}


\title{A $\mathsf{Q}$-tensor model for electrokinetics in nematic liquid crystals}


\author{O.~M.~Tovkach}
\email[]{otovkach@umass.edu}
\affiliation{Department of Mathematics, The University of Akron, Akron, OH 44325, USA. Currently at: Department of Physics, University of Massachusetts, Amherst, MA 01003}
\affiliation{Bogolyubov Institute for Theoretical Physics, NAS of Ukraine, Metrologichna 14-b, Kyiv 03680, Ukraine}
\author{Christopher~Conklin}
\email[]{conk0044@umn.edu}
\affiliation{School of Physics and Astronomy and Minnesota Supercomputing Institute, University of
Minnesota, Minneapolis, MN 55455, USA}
\author{M.~Carme~Calderer}
\email[]{calde014@umn.edu}
\affiliation{School of Mathematics, University of Minnesota, Minneapolis, MN 55455, USA}
\author{Dmitry~Golovaty}
\email[]{dmitry@uakron.edu}
\affiliation{Department of Mathematics, The University of Akron, Akron, OH 44325, USA}
\author{Oleg~D.~Lavrentovich}
\email[]{olavrent@kent.edu}
\affiliation{Liquid Crystal Institute, Kent State University, Kent, OH 44242, USA}
\author{Jorge~Vi\~nals}
\email[]{vinals@umn.edu}
\affiliation{School of Physics and Astronomy and Minnesota Supercomputing Institute, University of
Minnesota, Minneapolis, MN 55455, USA}
\author{Noel~J.~Walkington}
\email[]{noelw@andrew.cmu.edu}
\affiliation{Department of Mathematical Sciences, Carnegie Mellon University, Pittsburgh, PA 15213, USA}


\date{\today}

\begin{abstract}
We use a variational principle to derive a mathematical model for a nematic electrolyte in which the liquid crystalline component is described in terms of a second-rank order {parameter} tensor. The model extends the previously developed director-based theory and accounts for presence of disclinations and possible biaxiality. We verify the model by considering a simple but illustrative example of liquid crystal-enabled electro-osmotic flow around a stationary dielectric spherical particle placed at the center of a large cylindrical container filled with a nematic electrolyte. Assuming homeotropic anchoring of the nematic on the surface of the particle and uniform distribution of the director on the surface of the container, we consider two configurations with a disclination equatorial ring and with a hyperbolic hedgehog, respectively.  The computed electro-osmotic flows show a strong dependence on the director configurations and on the anisotropies of dielectric permittivity and electric conductivity of the nematic, characteristic of liquid crystal-enabled electrokinetics. Further, the simulations demonstrate space charge separation around the dielectric sphere, even in the case of isotropic permittivity and conductivity. {This is in agreement with the induced-charge electro-osmotic effect that occurs in an isotropic electrolyte when an applied field acts on the ionic charge it induces near a polarizable surface.}
\end{abstract} 


\maketitle

\section{Introduction}
Recent advancements in micro- and nanofluidics motivated a significant interest in electrokinetic phenomena, both from the theoretical and applied points of view \cite{Ramos, Morgan}. These phenomena occur in systems exhibiting spatial separation of charges that results from dissociation of polar chemical groups at the solid-fluid electrolyte interface and subsequent formation of an equilibrium electric double layer.  Application of the electric field causes electrokinetic flows with a characteristic velocity proportional to the applied field.  Besides this classical effect, there is a broad class of phenomena in which separation of charges is caused by the electric field itself \cite{Bazant}-\nocite{gamayunov1986pair,dukhin1986pair,murtsovkin1996nonlinear}\cite{squires2004induced}.  Since the induced charge is proportional to the applied field, the resulting flow velocities grow with the square of the field, $v\sim E^2$, \cite{Bazant}-\nocite{gamayunov1986pair,dukhin1986pair,gamayunov1986pair}\cite{squires2004induced}.  These phenomena, called collectively an {\it Induced-Charge Electrokinetics} (ICEK) \cite{Bazant}, are most often considered for ideally polarizable (conducting) solid particles. {The ICEK} velocity scale is $v^{metal}\sim\varepsilon_0\varepsilon_{medium}E^2a/\eta$, where $a$ is the radius of the colloid while $\varepsilon_0\varepsilon_{medium}$ and $\eta$ are the dielectric permittivity and viscosity of the electrolyte, respectively \cite{Bazant,murtsovkin1996nonlinear,squires2004induced}. If the particle is a solid dielectric with permittivity $\varepsilon_0\varepsilon_p$, the  ICE{K} flows are still present but with a much reduced velocity,  $v^{diel}\sim\varepsilon_0\varepsilon_pE^2\lambda_D/\eta$, where $\lambda_D$ is the Debye screening length \cite{murtsovkin1996nonlinear,squires2004induced}. In aqueous electrolytes $\lambda_D$ is typically much smaller than $a$ (tens of nanometers vs micrometers).

Another mechanism to achieve charge separation---even without a solid component---is to use an anisotropic fluid, such as a nematic liquid crystal, as an electrolyte \cite{Hernandez_1, Hernandez_2, Lavrentovich_Nature, Lazo_1, Lazo_2, Sasaki}. The anisotropy of the medium in presence of spatial gradients of the orientational order makes it possible to move charged ions to different locations.  The subsequent motion of the fluid induced by the electric field gives rise to nonlinear effects \cite{Lavrentovich_Nature}, called the Liquid-Crystal-Enabled electrokinetics (LCEK) \cite{Hernandez_1, Hernandez_2, Lavrentovich_Nature, Lazo_1, Lazo_2, Peng_pattern,we_pattern}.  Both the experiments and theoretical considerations demonstrate that the LCEK flow velocities are proportional to the square of the electric field \cite{Hernandez_1, Hernandez_2, Lavrentovich_Nature, Lazo_1, Lazo_2, Peng_pattern,we_pattern}.  Because the flow direction is independent of the field polarity, LCEK transport can be driven by an alternating current, a feature desired in technological applications. {Note that the term "electokinetics"---as applied to a small solid particle located in a fluid electrolyte---embraces two complementary phenomena. The first is {\it electrophoresis}, i.e., the motion of the particle with respect to the fluid under the action of a uniform electric field.  The second is {\it electro-osmosis}, the motion of a fluid electrolyte with respect to the particle that is immobilized (for example, glued to the substrate), also under the action of the externally applied uniform electric field.   This classification can be applied to both the ICEK and LCEK.  In particular, \cite{Lavrentovich_Nature,Lazo_1} describe electrophoresis of solid colloidal particles freely suspended in a nematic electrolyte, while \cite{Lazo_2} deals with a flow of the nematic electrolyte around solid particles that are glued to a substrate; in both cases, the characteristic velocities grow as $\sim E^2$. The first effect is called the Liquid-Crystal-Enabled Electrophoresis (LCEP) \cite{Lazo_1}, while the second is known as the Liquid-Crystal-Enabled Electro-osmosis (LCEO) \cite{Lazo_2}. In our work, we deal with the LCEO effect, considering an immobilized spherical particle in the nematic electrolyte.}

In this paper we derive a mathematical model for {electro-osmotic} flows in nematic liquid crystals, where the nematic component is described by the second-rank tensor order parameter, known as the $\mathsf{Q}$-tensor. The model generalizes our previous work that extended Ericksen-Leslie formalism \cite{Leslie_1, Leslie_2, Walkington} to nematic electrolytes, where we  established a system of governing equations from the local form of balance of linear and angular momentum within the framework of the director-based theory. An alternative derivation can be found in \cite{we_pattern}, where we arrived at the same system of equations in a more formal, but also more efficient manner following a variational formulation of nematodynamics, as proposed in \cite{Sonnet_tensor, Sonnet_dissipative}. Because the director models have a limited applicability {\cite{KleLa,BaZa}} in that they cannot model nematic biaxiality and topological defects---other than vortices---here we use the strategy in \cite{Sonnet_tensor, Sonnet_dissipative,we_pattern} to arrive at the appropriate $\mathsf{Q}$-tensor-based theory. 

As an illustrative example, we consider a stationary, relatively small (sub-micrometer) colloidal sphere that sets a perpendicular surface anchoring of the preferred orientation of the nematic. The director field around the particle is either of the quadrupolar type with an equatorial disclination loop \cite{kuksenok1996director} or  of dipolar symmetry, with {a  hyperbolic hedgehog point defect (strictly speaking, a small disclination ring; see, e.g., \cite{PhysRevLett.116.147801} and references therein.)} residing on one side of the sphere \cite{poulin1997novel}. Numerical simulations demonstrate electro-osmotic flows around these two configurations that are in qualitative agreement with the experimental data \cite{Lazo_2} but also highlight features characteristic for the {Induced-Charge Electro-osmotic (ICEO) flows} around a dielectric sphere in absence of materials anisotropies \cite{gamayunov1986pair,murtsovkin1996nonlinear,squires2004induced}. {Note that, since disclinations loops cannot be modeled within the framework of a director-based theory, this particular system is beyond the scope of the approach that we developed in \cite{we_pattern}.}

The paper is organized as follows. In Section \ref{s:1} we recall the principle of minimum energy dissipation and then use this principle in Section \ref{s:2} to derive the system of governing equations for our model. In Section \ref{s:3} we solve the governing system numerically to obtain the flow and charge patterns for electrokinetic flows around a stationary spherical  particle in a cylindrical column of a nematic electrolyte.

\section{Principle of minimum energy dissipation}
\label{s:1}

There is a variety of variational principles governing behavior of evolutionary systems \cite{Mielke}.
In classical mechanics, for instance, irreversible dynamics of a system can be described by means of a Rayleigh dissipation function $\mathcal{R}=\frac{1}{2}\xi_{ij}\dot{q}_i\dot{q}_j$ quadratic in generalized velocities $\dot{q}=(\dot{q}_1, ..., \dot{q}_M)$  (summation over repeated subscripts is implied hereafter).
The basic idea is to balance frictional and conservative forces in Lagrange's dynamical equations 
\begin{equation}\label{Lagrange_eq}
\frac{d}{dt}\frac{\partial \mathcal{L}}{\partial \dot{q}_m} -\frac{\partial \mathcal{L}}{\partial q_m} +\frac{\partial \mathcal{R}}{\partial \dot{q}_m} = 0,
\end{equation}
where $q=(q_1, ..., q_M)$ are generalized coordinates conjugated with the velocities $\dot{q}$ and $\mathcal{L}=\frac{1}{2}a_{ij}(q)\dot{q}_i\dot{q}_j-\mathcal{U}(q)$ is the Lagrangian of the system, defined as the difference between the kinetic energy $\frac{1}{2}a_{ij}(q)\dot{q}_i\dot{q}_j$ and the potential energy $\mathcal{U}(q)$. In what follows, we assume that the matrices $\left(\xi_{ij}\right)$ and $\left(a_{ij}\right)$ are symmetric.

Similarly to their non-dissipative counterparts,  Eqs.~\eqref{Lagrange_eq} can be recast into a variational problem as their solutions provide critical points of the functional 
$$\int_{\Omega} d^3r\left\{ \dot{\mathcal{E}}+\mathcal{R} \right\}$$
with respect to a special class of variations $\delta\dot{q}$ of the generalized velocities $\dot{q}$. 
Here $\Omega\subset\mathbb R^3$ is the region occupied by the system, $\mathcal{E}=\mathcal{L}+2\mathcal{U}$ is the total energy and the superimposed dot (as well as $\frac{d}{dt}$) denotes the total or material time derivative.
Unlike Hamilton's principle of stationary action, the current approach ``freezes'' both the configuration $q$ and the generalized forces $X_m:=\frac{d}{dt}\frac{\partial\mathcal{L}}{\partial \dot{q}_m}-\frac{\partial\mathcal{L}}{\partial q_m}$, $m=1,\ldots,M$ acting on the system at a given time. The state of the system is then varied by imposing arbitrary instantaneous variations $\delta\dot{q}$ of the velocities $\dot{q}$. Note that variations $\delta q$, $\delta\dot{q}$, and $\delta\ddot{q}$ are mutually independent except for the condition that the generalized forces $X_m,\ m=1,\ldots,M$ should remain unaltered \cite{Sonnet_book}.
Then, by using the product rule and relabeling, we indeed have
\begin{multline}
\frac{\delta}{\delta \dot{q}_m} \int_{\Omega} d^3r \left\{ \dot{\mathcal{E}} +\mathcal{R} \right\} =
\frac{\delta}{\delta \dot{q}_m} \int_{\Omega} d^3r \left\{ a_{ij}\ddot{q}_j\dot{q}_i +\frac{1}{2}\frac{\partial a_{ij}}{\partial q_k}\dot{q}_k\dot{q}_j\dot{q}_i +\frac{\partial \mathcal{U}}{\partial q_i}\dot{q}_i +\mathcal{R} \right\} \\
= \frac{\delta}{\delta \dot{q}_m} \int_{\Omega} d^3r \left\{ \left[\frac{d}{dt}\left( a_{ij}\dot{q}_j \right) -\frac{1}{2}\frac{\partial a_{kj}}{\partial q_i}\dot{q}_k\dot{q}_j +\frac{\partial \mathcal{U}}{\partial q_i}\right]\dot{q}_i +\mathcal{R} \right\} =\frac{\delta}{\delta \dot{q}_m} \int_{\Omega} d^3r \left\{ X_i\dot{q}_i +\mathcal{R} \right\} \\
=X_m+\frac{\partial \mathcal{R}}{\partial \dot{q}_m}=\frac{d}{dt} \frac{\partial \mathcal{L}}{\partial \dot{q}_m} -\frac{\partial \mathcal{L}}{\partial q_m} +\frac{\partial \mathcal{R}}{\partial \dot{q}_m},
\end{multline}
for every $m=1,\ldots,M$.
Hence, the Euler-Lagrange equations
\begin{equation}\label{Principle}
\frac{\delta}{\delta \dot{q}} \int_{\Omega} d^3r \left\{ \dot{\mathcal{E}} +\mathcal{R}\right\} = 0
\end{equation}
are identical to the generalized equations of motion \eqref{Lagrange_eq} and thus govern dynamics of a dissipative mechanical system. Since the conservative forces  are assumed to be fixed here and $\mathcal{R}$ is a positive-definite function, the equations \eqref{Principle} yield a minimum of energy dissipation \cite{Sonnet_tensor, Sonnet_dissipative}.
It is worth noting that for overdamped systems---where $\ddot{q}=0$---this principle of minimum energy dissipation is equivalent to the Onsager's variational approach \cite{Doi_PR}.  

\section{Nematic electrolyte}
\label{s:2}

In this section, we apply the principle \eqref{Principle} to a nematic electrolyte subject to an external electric field.
It was shown earlier that under an appropriate choice of the generalized velocities this framework is capable of reproducing the classical Ericksen-Leslie equations of nematodynamics \cite{Sonnet_tensor, Sonnet_dissipative}, {as well as the equations for ionic transport \cite{Xu}, and flow and sedimentation in colloidal suspensions \cite{Doi_PR}.} Below we demonstrate that it can be extended so as to take into account the presence of an ionic subsystem.

\subsection{Energy of the system}

{Consider a nematic liquid crystal that contains $N$ species of ions with valences $z^{\alpha}$ at concentrations $c^{\alpha}$, where $1\leq\alpha\leq N$. We assume that all ionic concentrations are small so that the resulting electrolyte solution is dilute. In LCEO experiments \cite{Lazo_2}, the concentration of ions is on the order of $10^{19}$ m$^{-3}$, which correspond to typical distances between isolated ions to be rather large, $\sim 0.5$ micrometer in absence of the electric field.} Then one can write the energy density of the ionic subsystem {as a sum} of the entropic and Coulombic contributions  
\begin{equation}\label{E_ion}
\mathcal{E}_{ion} = k_{B} \Theta \sum_{\alpha=1}^{N} c^{\alpha} \ln c^{\alpha} +\sum_{\alpha=1}^{N} e c^{\alpha} z^{\alpha} \Phi,
\end{equation}
where $k_B$ and $\Theta$ stand for the Boltzmann constant and the absolute temperature, respectively, $\Phi$ denotes the electric potential, and $e$ the elementary charge.
Under the action of the field, the ions move with {the effective} velocities $\mathbf{u}^{\alpha}$ which satisfy the continuity equations 
\begin{equation}\label{Charge_conservation}
\frac{\partial c^{\alpha}}{\partial t} + \nabla\cdot (c^{\alpha}\mathbf{u}^{\alpha}) = 0.
\end{equation}

Nematics themselves are anisotropic ordered fluids.
A typical nematic consists of elongated molecules whose local orientation can be described by a coarse-grained vector field $\mathbf{n}\equiv -\mathbf{n}$ with non-polar symmetry, the director. This unit-length vector field appropriately describes uniaxial nematic states with constant degree of orientational order  $S$.

In general, the degree of orientational order may not be constant, a nematic may contain disclinations, or be in a biaxial state (characterized by a spatially varying degree of biaxiality $P(\mathbf{r})$ and a set of not one, but two mutually orthogonal unit-length vector fields). Neither of these effects can be modeled within the framework of the standard director theory {\cite{KleLa,BaZa}}. The appropriate order parameter to characterize all available nematic states is a symmetric traceless second rank tensor $\mathsf{Q}$ with three, possibly different, eigenvalues.
In the uniaxial limit, two of the eigenvalues are equal so that
\begin{equation}\label{Q_tensor}
\mathsf{Q}_{ij} = S(n_in_j-\frac{1}{3}\delta_{ij}).
\end{equation}   
Then the free energy per a unit volume of a nematic liquid crystal can be written in the following form
\begin{equation}\label{E_LdG}
\mathcal{E}_{LdG} = -\frac{A}{2}\mathsf{Q}_{ij}\mathsf{Q}_{ij} +\frac{B}{3}\mathsf{Q}_{ij}\mathsf{Q}_{jk}\mathsf{Q}_{ki} +\frac{C}{4}(\mathsf{Q}_{ij}\mathsf{Q}_{ij})^2 +\frac{L}{2}(\partial_k\mathsf{Q}_{ij})(\partial_k\mathsf{Q}_{ij}),
\end{equation} 
where the first three terms represent the so-called Landau-de Gennes potential
\begin{equation}\label{E_LdGp}
\mathcal{E}_{LdG}^p = -\frac{A}{2}\mathsf{Q}_{ij}\mathsf{Q}_{ij} +\frac{B}{3}\mathsf{Q}_{ij}\mathsf{Q}_{jk}\mathsf{Q}_{ki} +\frac{C}{4}(\mathsf{Q}_{ij}\mathsf{Q}_{ij})^2,
\end{equation} 
given by an expansion of the free energy of the nematic in terms of the order parameter. The last term $\frac{L}{2}(\partial_k\mathsf{Q}_{ij})(\partial_k\mathsf{Q}_{ij})=\mathcal{E}_{LdG}^{e}$ in \eqref{E_LdG} accounts for elasticity of the liquid crystal with one elastic constant approximation being adopted from now on.

In order to take into account the interaction between the electric field $\mathbf{E}=-\nabla\Phi$ and the liquid crystal, we have to supplement the potential energy \eqref{E_LdG} of the nematic by 
\begin{equation}\label{E_E}
\mathcal{E}_{E} = -\frac{1}{2}\mathbf{D}\cdot\mathbf{E},
\end{equation}
where $\mathbf{D}$ denotes the electric displacement vector that satisfies
\begin{equation}\label{Maxwell_eq}
\nabla\cdot\mathbf{D}=\sum_{\alpha=1}^{N} e c^{\alpha} z^{\alpha}.
\end{equation}
It should be noted that care must be taken in dealing with the electric field in this problem.
The field is substantially nonlocal, that is, its changes can affect the system even if they occur outside the region $\Omega$ occupied by the system.
In order to avoid dealing with the field outside of $\Omega$, we assume that the system under investigation is surrounded by conductors that are held at a prescribed potential $\Phi_{\partial\Omega}$.
Then the electric field exists in $\Omega$ only, so that $D_i = \varepsilon_0\varepsilon_{ij}E_j$ where
\begin{equation}\label{Epsilon_ij}
\varepsilon_{ij} = \frac{1}{3}(\varepsilon_{\|}+2\varepsilon_{\perp})\delta_{ij} +\Delta\varepsilon\mathsf{Q}_{ij}
\end{equation}
with $\Delta\varepsilon=\varepsilon_{\|}-\varepsilon_{\perp}$, $\varepsilon_{\perp}$ and $\varepsilon_{\|}$ being dielectric permittivities perpendicular and along the director, respectively, measured in units of the vacuum permittivity $\varepsilon_0$.
Equation~\eqref{Epsilon_ij} can, in fact, used as an implicit phenomenological definition of the tensor order parameter $\mathsf{Q}$. 

Thus, neglecting inertia of molecular rotations $(\ddot{\mathsf{Q}}_{ij}=0)$, one can write the total energy per unit volume of the system in the form
\begin{equation}\label{Total_energy}
\mathcal{E}= \frac{1}{2}\rho v_i v_i +\mathcal{E}_{LdG} +\mathcal{E}_{E} +\mathcal{E}_{ion}
\end{equation}
with $\rho$ being the nematic mass density and $\mathbf{v}$ {the macroscopic velocity} of its flow which we assume to be incompressible, $\nabla\cdot\mathbf{v}=0$. {The incompressibility assumption is justified since a typical electrokinetic velocity is negligibly small compared to the speed of sound in nematics \cite{PhysRevLett.28.799}, corresponding to the Mach number $\sim 10^{-8}$. Note that the assumption that the nematic electrolyte solution is dilute allows us to think of $\rho$ and $\mathbf{v}$ as the density and the velocity of the nematic flow, respectively. Indeed, both of these quantities are defined as weighted volume averages of the velocities of the nematic and ionic constituents and the volume fraction of ions in the dilute solution is small.} 

\subsection{Dissipation function}

We require the dissipation function to be frame-indifferent, positive-definite and quadratic in the generalized velocities. 
Then, choosing $\mathbf{v}$ and $\dot{\mathsf{Q}}$ to be the generalized velocities, the dissipation function of a nematic liquid crystal $\mathcal{R}_{nem}$ has to be quadratic in $\mathbf{v}$ and $\dot{\mathsf{Q}}$.
This restriction, however, does not specify the dependence of the dissipation function on $\mathsf{Q}$ which, in general allows for a large number of nematic viscosity coefficients \cite{Sonnet_tensor}. Following \cite{Er91}, we reduce the number of these coefficients by restricting $\mathcal{R}_{nem}$ to the terms that are at most quadratic in the scalar order parameter $S$. Then
\begin{multline}
2\mathcal{R}_{nem} = \zeta_1\mathring{\mathsf{Q}}_{ij}\mathring{\mathsf{Q}}_{ji} +2\zeta_2\mathsf{A}_{ij}\mathring{\mathsf{Q}}_{ji} +2\zeta_3\mathsf{A}_{ij}\mathring{\mathsf{Q}}_{jk}\mathsf{Q}_{ki} +2\zeta_4\mathsf{A}_{ij}\mathsf{A}_{jk}\mathsf{Q}_{ki} +\zeta_5\mathsf{A}_{ij}\mathsf{A}_{jk}\mathsf{Q}_{kl}\mathsf{Q}_{li}\\
 +\zeta_6\left(\mathsf{A}_{ij}\mathsf{Q}_{ji}\right)^2 +\zeta_7\mathsf{A}_{ij}\mathsf{A}_{ji}\mathsf{Q}_{kl}\mathsf{Q}_{lk} +\zeta_8\mathsf{A}_{ij}\mathsf{A}_{ji},
 \label{eq:stuff}
\end{multline}
where $\mathsf{A}_{ij}=\frac{1}{2}(\partial_j v_i +\partial_i v_j)$ represents the symmetric part of the velocity gradient and $\mathring{\mathsf{Q}}_{ij}=\dot{\mathsf{Q}}_{ij}-\mathsf{W}_{ik}\mathsf{Q}_{kj}-\mathsf{W}_{jk}\mathsf{Q}_{ki}$, with $\mathsf{W}_{ij}=\frac{1}{2}(\partial_jv_i-\partial_iv_j)$, gives the rate of the $\mathsf{Q}$-tensor change relative to a flow vorticity \cite{Sonnet_tensor}.
Inserting the uniaxial representation \eqref{Q_tensor} of the tensorial order parameter $\mathsf{Q}$ in \eqref{eq:stuff} and taking into account that $\mathring{n}_i=\dot{n}_i-\mathsf{W}_{ij}n_j$ and $\dot{S}=0$, the dissipation function takes the form
\begin{equation}
2\mathcal{R}_{nem}^{(\mathbf{n})} = (\alpha_3-\alpha_2) \mathring{n_i}^2 +2(\alpha_5-\alpha_6) \mathring{n_i} \mathsf{A}_{ij}n_j +(\alpha_5+\alpha_6)(\mathsf{A}_{ij}n_j)^2 +\alpha_4(\mathsf{A}_{ij})^2+\alpha_1(n_i \mathsf{A}_{ij} n_j)^2,
\end{equation}
 when written in terms of the director $\mathbf{n}$. Now one can relate the viscosities $\zeta_i$ to the Leslie's viscosities $\alpha_j$ \cite{DK84}:
\begin{equation}\label{Viscosities}
\begin{split}
\alpha_3-\alpha_2 =& 2S^2\zeta_1, \qquad\qquad
\alpha_6-\alpha_5 = 2S\zeta_2 +\frac{1}{3}S^2\zeta_3,\\
\alpha_1 =& S^2\zeta_6, \qquad\qquad
\alpha_5+\alpha_6 = S\zeta_4 +\frac{1}{2}S^2\zeta_5,\\
\alpha_4 =& \zeta_8 -\frac{1}{3}S\zeta_4 +\frac{1}{3}S^2\left(\frac{1}{3}\zeta_5+2\zeta_7\right).
\end{split}
\end{equation}
It follows from \eqref{Viscosities} that the viscosities $\zeta_3$, $\zeta_5$, and $\zeta_7$ are higher-order corrections to the Leslie's viscosities in terms of the scalar order parameter $S$. 
Thus, one can set $\zeta_3=\zeta_5=\zeta_7=0$ and arrive at a simpler form of the dissipation function
\begin{equation}\label{R_nem}
2\mathcal{R}_{nem} = \zeta_1\mathring{\mathsf{Q}}_{ij}\mathring{\mathsf{Q}}_{ji} +2\zeta_2\mathsf{A}_{ij}\mathring{\mathsf{Q}}_{ji} +2\zeta_4\mathsf{A}_{ij}\mathsf{A}_{jk}\mathsf{Q}_{ki} +\zeta_6\left(\mathsf{A}_{ij}\mathsf{Q}_{ji}\right)^2 +\zeta_8\mathsf{A}_{ij}\mathsf{A}_{ji},
\end{equation}  
which involves only five nematic viscosities. {As we will demonstrate below, this choice of $\mathcal{R}_{nem}$ results in the expression for viscous stress identical to that derived in \cite{Ping_Sheng}.}

For the nematic electrolyte, we also need to incorporate dissipation due to the motion of ions. Taking into account that the mobilities of ions along and perpendicular to the director $\mathbf{n}$ are different and treating $\mathbf{u}^{\alpha}$ with $1\leq\alpha\leq N$ as the generalized velocities, the contribution of ions to dissipation is given by \cite{Carme}
\begin{equation}\label{R_ion}
2\mathcal{R}_{ion} = k_B\Theta \sum_{\alpha=1}^{N} c^{\alpha}(\mathsf{D}_{ij}^{\alpha})^{-1}(u_i^{\alpha}-v_i)(u_j^{\alpha}-v_j).
\end{equation}
Here the diffusion matrix $\mathsf{D}_{ij}^{\alpha}$ accounts for the anisotropy of the liquid crystal electrolyte. {The expression \eqref{R_ion} is a direct generalization of the dissipation function for ordinary colloidal solutions \cite{Doi_PR}. }

Thus, the total energy dissipation rate in the nematic electrolyte is equal to the sum $\mathcal{R}=\mathcal{R}_{nem}+\mathcal{R}_{ion}$ with $\mathcal{R}_{nem}$ as specified in \eqref{R_nem}.

\subsection{Governing equations}

Once the energy $\mathcal{E}$, the dissipation $\mathcal{R}$, and the generalized velocities of the system are specified, we are in a position to derive equations describing electro-osmotic flows in nematics.
The equations are implicitly given by
\begin{equation}\label{EL_eq}
\begin{split}
\frac{\delta}{\delta \mathbf{v}}\int_{\Omega} d^3r \left\{ \dot{\mathcal{E}}+\mathcal{R} -p^{\prime} (\partial_i v_i) -\Lambda \mathsf{Q}_{ii}  \right\}=0, \\
\frac{\delta}{\delta \dot{\mathsf{Q}}}\int_{\Omega} d^3r \left\{ \dot{\mathcal{E}}+\mathcal{R} -p^{\prime} (\partial_i v_i) -\Lambda \mathsf{Q}_{ii}  \right\}=0, \\
\frac{\delta}{\delta \mathbf{u}^{\alpha}}\int_{\Omega} d^3r \left\{ \dot{\mathcal{E}}+\mathcal{R} -p^{\prime} (\partial_i v_i) -\Lambda \mathsf{Q}_{ii}  \right\}=0,
\end{split}
\end{equation}
where the two Lagrange multipliers, $p^{\prime}$ and $\Lambda$, are associated with the flow incompressibility and the tracelessness of the tensor order parameter, respectively.

But before deriving the explicit form of \eqref{EL_eq}, let us specify the boundary conditions for our problem.
Although one can simply use the natural boundary conditions that follow from the principle of minimum energy dissipation \eqref{Principle}, here we impose Dirichlet conditions on $\partial\Omega$. In particular,
\begin{equation}
\label{eq:bc}
\mathbf{v}=0,\quad \dot{\mathsf{Q}}=0,\quad \mathbf{u}^{\alpha}=0 \quad \text{on } \partial\Omega.
\end{equation}
This choice of boundary conditions slightly simplifies further consideration and corresponds to a majority of experimental setups.

Next, we calculate the rate of change of the energy in Eq.~\eqref{EL_eq}: we start by computing
\begin{multline}\label{E_nem_dot}
\frac{d}{d t} \int_{\Omega} d^3r \left\{ \frac{1}{2} \rho \mathbf{v}^2 +\mathcal{E}_{LdG}\left(\mathsf{Q}, \nabla\mathsf{Q}\right) \right\} =  \\
= \int_{\Omega} d^3r \left\{\left[ \rho\dot{v_l} +\partial_k\left(\frac{\partial \mathcal{E}_{LdG}}{\partial (\partial_k \mathsf{Q}_{ij})}(\partial_l \mathsf{Q}_{ij})\right) \right] v_l +\left[\frac{\partial \mathcal{E}_{LdG}}{\partial \mathsf{Q}_{ij}} -\partial_k\left(\frac{\partial \mathcal{E}_{LdG}}{\partial (\partial_k \mathsf{Q}_{ij})}\right) \right]\dot{\mathsf{Q}}_{ij}   \right\},
\end{multline}
and
\begin{equation}
\frac{d}{dt} \int_{\Omega} d^3r \mathcal{E}_{E}(\mathsf{Q}, \nabla\Phi) = 
\int_{\Omega} d^3r \left\{ \frac{\partial \mathcal{E}_{E}}{\partial \mathsf{Q}_{ij}}\dot{\mathsf{Q}}_{ij} +\frac{\partial \mathcal{E}_{E}}{\partial (\partial_i \Phi)}(\partial_i \dot{\Phi}) -\frac{\partial \mathcal{E}_{E}}{\partial (\partial_i \Phi)}(\partial_k \Phi)(\partial_i v_k)\right\},
\end{equation}
with help of the identity $\dot{(\partial_k\mathsf{Q}_{ij})}=\partial_k\dot{\mathsf{Q}}_{ij} -(\partial_k v_l)(\partial_l \mathsf{Q}_{ij})$.

Recall that $$\mathcal{E}_{E} = -\varepsilon_0 (\bar{\varepsilon}\delta_{ij}+\Delta\varepsilon \mathsf{Q}_{ij})(\partial_i \Phi) (\partial_j \Phi)/2,$$ where $\bar{\varepsilon}=(\varepsilon_{\|}+2\varepsilon_{\perp})/3$. Then
\begin{equation}
\frac{\partial \mathcal{E}_E}{\partial \mathsf{Q}_{ij}} = -\frac{1}{2}\varepsilon_0 \Delta\varepsilon (\partial_i \Phi) (\partial_j \Phi)\quad\mathrm{and}\quad\frac{\partial \mathcal{E}_E}{\partial (\partial_i \Phi)} = -\varepsilon_0 \varepsilon_{ij} (\partial_j \Phi).
\end{equation}
Hence
\begin{multline}\label{E_field_dot}
\frac{d}{dt} \int_{\Omega} d^3r \mathcal{E}_{E}(\mathsf{Q}, \nabla\Phi) = \\
= \int_{\Omega} d^3r \left\{ -\frac{1}{2}\varepsilon_0\Delta\varepsilon E_i E_j \dot{\mathsf{Q}}_{ij} -(\partial_i D_i) \dot{\Phi} -\partial_i (\varepsilon_0\varepsilon_{ij} E_j E_k) v_k  \right\} +\int_{\partial\Omega} d^2r \left\{ (\nu_i \varepsilon_0\varepsilon_{ij}E_j) \dot{\Phi} \right\}.
\end{multline}
On a conductor-dielectric interface, the normal component of the displacement, $D_i\nu_i$, is given by the surface charge density $\sigma$. It follows from \eqref{eq:bc} and the definition of a material derivative that the surface integral in \eqref{E_field_dot}  can be written as
\begin{equation}
\int_{\partial\Omega} d^2r \left\{ (\nu_i \varepsilon_0\varepsilon_{ij}E_j) \dot{\Phi} \right\} = \int_{\partial\Omega} d^2r\, D_i\nu_i  \frac{\partial\Phi}{\partial t} =\int_{\partial\Omega} d^2r\, \sigma \frac{\partial \Phi}{\partial t}.
\end{equation}
This integral gives the power spent by charges located at $\partial\Omega$ and can be omitted when $\Phi_{\partial\Omega}$ varies slowly compared to the timescales of the dynamics associated with $\mathbf{v}$, $\mathbf{u}^{\alpha}$ and $\dot{\mathsf{Q}}$. 

For the ionic subsystem, we have
\begin{equation}\label{E_ion_dot}
\frac{d}{dt}\int_{\Omega} d^3r \mathcal{E}_{ion}(c^{\alpha}, \Phi)=\\
\int_{\Omega} d^3r \sum_{\alpha=1}^{N} \left\{(\partial_i \mu^{\alpha}) c^{\alpha} (u_i^{\alpha} -v_i) +e c^{\alpha}z^{\alpha}\dot{\Phi} -\mu^{\alpha}c^{\alpha}(\partial_i v_i) \right\},
\end{equation}
where $\mu^{\alpha} = \frac{\partial \mathcal{E}_{ion}}{\partial c^{\alpha}} = k_{B}\Theta (\ln c^{\alpha} +1) +e z^{\alpha} \Phi$ is the chemical potential of the $\alpha$-th ion species \cite{Eisenberg}.

Note that $\dot{\mathcal{E}}_{ion}$ includes the term $\sum_{\alpha}ec^{\alpha}z^{\alpha}\dot{\Phi}$ whereas $\dot{\mathcal{E}}_{E}$ contains $-(\partial_i D_i)\dot{\Phi}$; these terms cancel out when combined together in the expression for the total power $\dot{\mathcal{E}}$. This is due to the fact that the electric field obeys the Maxwell's equation \eqref{Maxwell_eq}. 

We could have instead obtained the same equation \eqref{Maxwell_eq} for $\mathbf{D}$ from \eqref{Principle}, if we chose to treat $\dot{\Phi}$ as a generalized velocity. Then
\begin{equation}
\frac{\delta}{\delta \dot{\Phi}}\int_{\Omega} d^3r \left\{ \dot{\mathcal{E}}+\mathcal{R} -p^{\prime} (\partial_i v_i) -\Lambda \mathsf{Q}_{ii}  \right\}=-\partial_i D_i +\sum_{\alpha=1}^{N} ec^{\alpha}z^{\alpha} = 0. 
\end{equation}
Since the present framework deals with the energy of the entire system this derivation properly addresses the nonlocality of the field.

The variational derivatives of the total dissipation function $\mathcal{R}$ are given by
\begin{eqnarray}\label{dR}
&\frac{\delta}{\delta \dot{\mathsf{Q}}_{ij}} \int_{\Omega}d^3r\mathcal{R}= \frac{\partial \mathcal{R}_{nem}}{\partial \mathring{\mathsf{Q}}_{ij}} = \zeta_1\mathring{\mathsf{Q}}_{ij} +\zeta_2 \mathsf{A}_{ij},& \label{dRdn}\\
&\frac{\delta}{\delta u_i^{\alpha}} \int_{\Omega}d^3r\mathcal{R} = k_B\Theta c^{\alpha}(\mathsf{D}_{ij}^{\alpha})^{-1}(u_j^{\alpha}-v_j),&\label{dRdu}\\
&\frac{\delta}{\delta v_i} \int_{\Omega}d^3r\mathcal{R}= \frac{\delta}{\delta v_i}\int_{\Omega}d^3r\mathcal{R}_{nem} - k_B\Theta \sum_{\alpha=1}^{N}c^{\alpha}(\mathsf{D}_{ij}^{\alpha})^{-1}(u_j^{\alpha}-v_j).&\label{dRdv}
\end{eqnarray}
Using the explicit form \eqref{R_nem} of $\mathcal{R}_{nem}$ and the chain rule
$$\frac{\partial}{\partial(\partial_j v_i)} = \frac{\partial}{\partial\mathsf{A}_{ij}} +\mathsf{Q}_{ki}\frac{\partial}{\partial\mathring{\mathsf{Q}}_{jk}} -\mathsf{Q}_{kj}\frac{\partial}{\partial\mathring{\mathsf{Q}}_{ik}},$$
we obtain that $$\frac{\delta}{\delta v_i}\int_{\Omega}d^3r\mathcal{R}_{nem} = -\partial_j\mathsf{T}_{ij}^V,$$ where the viscous stress tensor 
\begin{multline}\label{Viscous_stress}
\mathsf{T}_{ij}^V = \zeta_1 \left(\mathring{\mathsf{Q}}_{jk}\mathsf{Q}_{ki} -\mathring{\mathsf{Q}}_{ik}\mathsf{Q}_{kj}\right) +\zeta_2\mathring{\mathsf{Q}}_{ij} +(\zeta_4+\zeta_2)\mathsf{A}_{jk}\mathsf{Q}_{ki}+\\
+(\zeta_4-\zeta_2)\mathsf{A}_{ik}\mathsf{Q}_{kj} +\zeta_6 \left(\mathsf{A}_{kl}\mathsf{Q}_{lk}\right)\mathsf{Q}_{ij} +\zeta_8\mathsf{A}_{ij} 
\end{multline}
is identical to that suggested in \cite{Ping_Sheng}.

Thus, it follows from \eqref{E_ion_dot} and \eqref{dRdu} that
\begin{equation}\label{Nernst0_eq}
\frac{\delta}{\delta u_i^{\alpha}}\int_{\Omega} d^3r \left\{ \dot{\mathcal{E}}+\mathcal{R} -p^{\prime} (\partial_i v_i) -\Lambda n_i\dot{n_i}  \right\}= c^{\alpha}\left(\partial_i \mu^{\alpha} +k_B\Theta (\mathsf{D}_{ij}^{\alpha})^{-1}(u_j^{\alpha}-v_j)\right) = 0.
\end{equation}
Combining this with the continuity equation \eqref{Charge_conservation}, we arrive at
\begin{equation}\label{Nernst_eq}
\frac{\partial c^{\alpha}}{\partial t} +\partial_j \left[ c^{\alpha}v_j -\frac{c^{\alpha}}{k_B\Theta}\mathsf{D}_{ij}^{\alpha}(\partial_i \mu^{\alpha}) \right] = 0.
\end{equation}
Likewise, equations \eqref{E_nem_dot}, \eqref{E_field_dot} and \eqref{dRdn} yield
\begin{multline}\label{Leslie_eq}
\frac{\delta}{\delta \dot{\mathsf{Q}}_{ij}}\int_{\Omega} d^3r \left\{ \dot{\mathcal{E}}+\mathcal{R} -p^{\prime} (\partial_i v_i) -\Lambda \mathsf{Q}_{ii}  \right\}= \\
= \frac{\partial \mathcal{E}_{LdG}}{\partial \mathsf{Q}_{ij}} -\partial_k \left[ \frac{\partial \mathcal{E}_{LdG}}{\partial(\partial_k \mathsf{Q}_{ij})} \right] -\Lambda \delta_{ij} -\frac{1}{2}\varepsilon_0\Delta\varepsilon E_i E_j +\zeta_1\mathring{\mathsf{Q}}_{ij} +\zeta_2 \mathsf{A}_{ij} =0.
\end{multline}
Finally, combining \eqref{E_nem_dot}, \eqref{E_field_dot}, \eqref{E_ion_dot}, \eqref{dRdv} and \eqref{Nernst0_eq} we arrive at
\begin{multline}\label{Navier0_eq}
\frac{\delta}{\delta v_i}\int_{\Omega} d^3r \left\{ \dot{\mathcal{E}}+\mathcal{R} -p^{\prime} (\partial_i v_i) -\Lambda \mathsf{Q}_{ii}  \right\}= \\
= \rho \dot{v_i} +\partial_k \left[ \frac{\partial \mathcal{E}_{LdG}}{\partial(\partial_k \mathsf{Q}_{mn})}(\partial_i \mathsf{Q}_{mn}) -\mathsf{T}_{ik}^V -\varepsilon_0\varepsilon_{kj}E_j E_i \right] +\partial_i p^{\prime} +\partial_i\left[\sum_{\alpha=1}^{N}c^{\alpha}\mu^{\alpha}\right] =0.
\end{multline}
The sum $p^{\prime}+\sum_{\alpha}c^{\alpha}\mu^{\alpha}$ can be defined as the total pressure $p$, thus yielding an alternative form 
\begin{equation}\label{Navier_eq}
\rho \dot{v_i} +\partial_k \left[ \frac{\partial \mathcal{E}_{LdG}}{\partial(\partial_k \mathsf{Q}_{mn})}(\partial_i \mathsf{Q}_{mn}) +p\delta_{ik} -\mathsf{T}_{ik}^V -\varepsilon_0\varepsilon_{kj}E_j E_i \right] =0
\end{equation}
of \eqref{Navier0_eq}.
Equations \eqref{Maxwell_eq}, \eqref{Nernst_eq}, \eqref{Leslie_eq} and \eqref{Navier_eq} along with 
the definition of the chemical potential
\begin{equation}
\mu^{\alpha} = \frac{\partial \mathcal{E}_{ion}}{\partial c^{\alpha}} = k_{B}\Theta (\ln c^{\alpha} +1) +e z^{\alpha} \Phi
\end{equation}
and constraints $\nabla\cdot\mathbf{v}=0$,  $\mathsf{Q}_{ii}=0$ constitute the full set of equations governing electro-osmosis in nematic liquid crystals, which can be written in the following invariant form 
\begin{equation}\label{The_System}
\begin{cases}
\frac{\partial c^{\alpha}}{\partial t} +\text{div} \left[ c^{\alpha}\mathbf{v} -\frac{c^{\alpha}}{k_B\Theta}\mathsf{D}^{\alpha}(\nabla \mu^{\alpha}) \right] = 0,\\
\frac{\partial \mathcal{E}_{LdG}}{\partial \mathsf{Q}} -\text{div} \left[ \frac{\partial \mathcal{E}_{LdG}}{\partial(\nabla \mathsf{Q})} \right] -\Lambda \mathsf{I} -\frac{1}{2}\varepsilon_0\Delta\varepsilon \mathbf{E}\otimes\mathbf{E} +\zeta_1\mathring{\mathsf{Q}} +\zeta_2 \mathsf{A} =0,\\
\rho \dot{\mathbf{v}} +\text{div} \left[ -\mathsf{T}^{\text{el}} +p\mathsf{I} -\mathsf{T}^V -\varepsilon_0 \mathbf{E}\otimes \hat{\varepsilon}\mathbf{E} \right] =0,\\
\text{div} \left[ \frac{1}{3}\left(\varepsilon_{\|}+2\varepsilon_{\perp}\right)\mathbf{E}  +\Delta\varepsilon \mathsf{Q} \mathbf{E}\right] = \frac{e}{\varepsilon_0}\sum_{\alpha=1}^{N}c^{\alpha}z^{\alpha},\\
\mu^{\alpha} = k_{B}\Theta (\ln c^{\alpha} +1) +e z^{\alpha} \Phi,\\
\text{div } \mathbf{v} =0,\\
\text{Tr } \mathsf{Q} =0,
\end{cases}
\end{equation}
where the elastic stress tensor $\mathsf{T}^{\text{el}}=-\dfrac{\partial \mathcal{E}_{LdG}}{\partial(\partial_k \mathsf{Q}_{mn})}(\partial_i \mathsf{Q}_{mn}) \,\mathbf{e}_i\otimes\mathbf{e}_{k}$, the dielectric tensor $\hat{\varepsilon}=\varepsilon_{ij}\,\mathbf{e}_i\otimes\mathbf{e}_j$ and $\mathsf{I}$ is the identity tensor.

\section{electro-osmotic flow around a spherical particle}
\label{s:3}

In this section, we consider a simple but illustrative example of liquid crystal-enabled electro-osmotic flow (LCEO) around an immobilized spherical particle placed at the center of a large cylindrical domain filled with a nematic electrolyte.
Recently, a similar problem in a rectangular container was experimentally examined in \cite{Lazo_2}.
Despite the difference in geometry, the physical mechanism of LCEO is essentially the same in both cases.
The colloidal inclusion distorts the otherwise uniform ordering of the liquid crystal molecules, inducing spatial variations of the order tensor $\mathsf{Q}$ field.
In the presence of an electric field, inhomogeneities of $\mathsf{Q}$, along with the anisotropy of dielectric permittivity and conductivity of the liquid crystal give rise to spatial separation of electric charges present in the system. This field-induced charging of distorted regions of the nematic electrolyte is a distinctive feature of LCEO, which consequently yields {electro-osmotic} flow with the velocity quadratic in the electric field.
The profile of the flow, as will be seen below, depends on the symmetry of the tensor field $\mathsf{Q}$ as well as on anisotropies of ionic conductivities and the dielectric permittivity of the nematic.

Let us consider a micron-sized spherical colloidal particle suspended in a nematic electrolyte subject to a uniform electric field $\mathbf{E}=(0, 0, -E)$.
For the sake of simplicity, assume that the ionic subsystem consists of two species with valences $z^+=1$ and $z^-=-1$ and concentrations $c^{+}$ and $c^{-}$, respectively. We assume equal mobility matrices
\[\mathsf{D}^{+}=\mathsf{D}^{-}=\mathsf{D}_{ij}=\bar{\mathsf{D}}(\bar{\lambda}_{\sigma}\delta_{ij} +(\lambda_{\sigma}-1)\mathsf{Q}_{ij})\,\mathbf{e}_i\otimes\mathbf{e}_j,\] 
where $\left\{\mathbf{e}_i\right\}_{i=1,2,3}$ is a set of mutually orthonormal vectors in $\mathbb{R}^3$ and $\lambda_{\sigma}=\sigma_{\|}/\sigma_{\perp}>0$ denotes the ratio of the conductivity along and perpendicular to the nematic director, respectively; $\bar{\lambda}_{\sigma}=\frac{1}{3}(\lambda_{\sigma}+2)$ and $\bar{\mathsf{D}}>0$.

For further analysis of the system of governing equations \eqref{The_System}, it is convenient to introduce nondimensional variables
\begin{equation}
\begin{split}
\tilde{\mathbf{r}}=\frac{\mathbf{r}}{a}, \qquad \tilde{t}=\frac{t}{\bar{v}}, \qquad \tilde{\Phi}=\frac{\Phi}{Ea},\qquad \tilde{c}^{\pm}=\frac{c^{\pm}}{\bar{c}}, \tilde{\mathbf{v}}=\frac{\mathbf{v}}{\bar{v}}, \\ \tilde{p}=\frac{p}{\bar{p}}, \qquad \tilde{\mathsf{D}}_{ij}=\frac{\mathsf{D}_{ij}}{\bar{\mathsf{D}}}, \qquad \tilde{\mathsf{T}}^V_{ij}=\mathsf{T}^V_{ij}\frac{a}{\zeta_8 \bar{v}},
\end{split}
\end{equation}
where $a$ is the radius of the particle and $\bar{x}$ denotes the characteristic value of $x$.
Then omitting the tildes for notational simplicity, one can rewrite the system \eqref{The_System} in the following nondimensional form
\begin{equation}\label{The_System_nondim}
\begin{cases}
\text{Pe}\left(\dfrac{\partial c^{\pm}}{\partial t} +\text{div} \left[ c^{\alpha}\mathbf{v} \right]\right) -\text{div} \left[ \mathsf{D}\left(\nabla c^{\pm} \mp c^{\pm} G \mathbf{E} \right) \right] = 0,\\
\dfrac{\partial \mathcal{E}_{LdG}}{\partial \mathsf{Q}} -\text{div} \left[ \dfrac{\partial \mathcal{E}_{LdG}}{\partial(\nabla \mathsf{Q})} \right] -\Lambda \mathsf{I} -\dfrac{1}{2}\dfrac{a^2}{\xi_E^2} \mathbf{E}\otimes\mathbf{E} +\text{Er}\left(\dfrac{\zeta_1}{\zeta_8} \mathring{\mathsf{Q}} +\dfrac{\zeta_2}{\zeta_8} \mathsf{A}\right) =0,\\
\text{Re}\, \dot{\mathbf{v}} +\text{div} \left[ -\dfrac{1}{\text{Er}}\mathsf{T}^{\text{el}} +p\mathsf{I} -\mathsf{T}^V -\mathbf{E}\otimes\dfrac{\hat{\varepsilon}}{\varepsilon_{\perp}}\mathbf{E} \right] =0,\\
\text{div} \left[ \dfrac{1}{3}\left(\lambda_{\varepsilon}+2\right)\mathbf{E}  +\left(\lambda_{\varepsilon}-1\right) \mathsf{Q} \mathbf{E}\right] = B\left( c^+-c^- \right),\\
\text{div } \mathbf{v} =0,\\
\text{Tr } \mathsf{Q} =0,
\end{cases}
\end{equation}
which implies $\bar{p} = \dfrac{\zeta_8 \bar{v}}{a}$ and $\bar{v} = \dfrac{\varepsilon_0\varepsilon_{\perp}aE^2}{\zeta_8}$, and where the nondimensional parameters
\begin{equation}
\begin{split}
\text{Pe}=\frac{\bar{v}a}{\bar{\mathsf{D}}}, \qquad \text{Er}=\frac{\zeta_8 \bar{v} a}{L}, \qquad \frac{a^2}{\xi_E^2}=\frac{\varepsilon_0\Delta\varepsilon E^2 a^2}{L}, \\ \text{Re}=\frac{\rho\bar{v}a}{\zeta_8}, \qquad B=\frac{e\bar{c}a}{\varepsilon_0\varepsilon_{\perp}E}, \qquad G=\frac{eaE}{k_{B}\Theta}
\end{split}
\end{equation}
along with $\lambda_{\varepsilon}=\varepsilon_{\|}/\varepsilon_{\perp}$ are introduced. Here $\xi_E=\sqrt{L/(\varepsilon_0|\Delta\varepsilon|E^2)}$ is the electric coherence length. We consider the colloidal sphere to be relatively small, $a\approx\,\upmu$m; the rest of the parameters are close to the ones used in typical experiments on LCEO: $\rho\approx 1$~g/cm$^3$, $\Delta\varepsilon\approx 10$, $\varepsilon_{\perp}\approx 10$, $L\approx 10$~pN, $\bar{\mathsf{D}}\approx 5\cdot 10^{-11}$~m$^2$/s, $\zeta_8\approx 0.1$~Pa$\cdot$s, $\bar{c}=10^{19}$~m$^{-3}$, $E\approx 40$~mV/$\upmu$m, and $\Theta=293$~K. Then the nondimensional parameters have the following values
\begin{equation}
\text{Pe}\approx 0.03, \qquad \text{Er}\approx 0.01, \qquad \frac{a^2}{\xi_E^2}\approx 0.01, \qquad \text{Re}\approx 1\cdot10^{-8}, \qquad B\approx 0.45, \qquad G\approx 1.6.
\end{equation}
Smallness of the first three characteristic numbers is of particular importance in what follows. Since diffusive transport of ions prevails over advective transport (the Peclet number $\text{Pe}\ll 1$) and the elasticity of the liquid crystal dominates over its viscosity (the Ericksen number $\text{Er}\ll 1$), the order parameter $\mathsf{Q}$ and the concentrations of ions $c^+$ and $c^-$ are not significantly affected by the liquid crystal flow.
Moreover, due to the small ratio of the particle radius $a$ to the electric coherence length $\xi_E$, we can also neglect the influence of the electric field on the molecular alignment. 

{Among the parameters listed above, only the radius $a$ of the sphere has a value that is different from what was used in the experiment in \cite{Lazo_2}, where $a=1\,\upmu$m in simulations vs. $a=25\,\upmu$m in the experiment. This departure is motivated by the two closely related reasons, (i) by the fact that small particles in a large nematic domain can feature both dipolar director field with a hyperbolic hedgehog and a quadrupolar director with an equatorial disclination ring and (ii) by the fact that the model developed in our work allows us to describe the LCEO effects in presence of the disclination rings which are naturally stable around the small spheres. As discussed below, the relative stability of the two director geometries around a small sphere can be tuned by slightly adjusting the size of the particle.  This allows us to compare the electro-osmotic flow patterns for the two different symmetries of director distortions while keeping the physical parameters close to each other in the two cases.  As the particles become bigger, the hedgehog configuration in a large domain becomes progressively more stable, while the ring configuration needs to be supported either by an external field or by strong confinement \cite{PhysRevLett.85.4719}. In the experiments \cite{Lazo_2}, the comparison between the hedgehog and ring configuration was made possible by placing the spheres into a shallow cell with the thickness that is only slightly larger than the diameter of the spheres.  Proximity of bounding walls complicates the numerical analysis of the flows and to some extent masks the difference caused by the different symmetry of the director field near the surface of the spheres. To avoid the complications associated with the strong confinement, in what follows we analyze the case of the small particles. 

A significant computational simplification associated with choosing the particle to be small results from the decoupling of the equations in \eqref{The_System_nondim}. Note that in \cite{Lazo_2}, for a particle of radius $25\,\upmu$m, the experimentally observed velocity of propagation was $~4\,\upmu$m/s, which corresponds to $\text{Er}=O(1)$. The system \eqref{The_System} can still be solved numerically in this situation, but at a significantly higher computational cost since the equations remain fully coupled.} 

Thus, the system of equations \eqref{The_System_nondim} can be solved in three consecutive steps.
First, we find the alignment tensor $\mathsf{Q}$ from  
\begin{equation}\label{Eq_for_Q}
\begin{cases}
\dfrac{\partial \mathcal{E}_{LdG}}{\partial \mathsf{Q}} -\text{div} \left[ \dfrac{\partial \mathcal{E}_{LdG}}{\partial(\nabla \mathsf{Q})} \right] -\Lambda \mathsf{I} =0,\\
\text{Tr }\mathsf{Q}=0,
\end{cases}
\end{equation}
then calculate the concentrations $c^{\pm}(\mathbf{r})$ and the electric field $\mathbf{E}=-\nabla\Phi$ given by
\begin{equation}\label{Eq_for_E}
\begin{cases}
\text{div} \left[ \mathsf{D}\left(\nabla c^{\pm} \mp c^{\pm} G \mathbf{E} \right) \right] = 0,\\
\text{div} \left[ \frac{1}{3}\left(\lambda_{\varepsilon}+2\right)\mathbf{E}  +\left(\lambda_{\varepsilon}-1\right) \mathsf{Q}\mathbf{E}\right] = B\left( c^+-c^- \right),
\end{cases}
\end{equation}
and finally, solve 
\begin{equation}\label{Eq_for_v}
\begin{cases}
\text{div} \left[ -\dfrac{1}{\text{Er}} \mathsf{T}^{\text{el}} +p\mathsf{I} -\mathsf{T}^V -\frac{1}{\varepsilon_{\perp}} \mathbf{E}\otimes\hat{\varepsilon}\mathbf{E} \right] =0,\\
\text{div } \mathbf{v} = 0
\end{cases}
\end{equation}
for the pressure $p(\mathbf{r})$ and the velocity field $\mathbf{v}(\mathbf{r})$.

\begin{figure}
\begin{center}
\includegraphics[width=.36\textwidth]{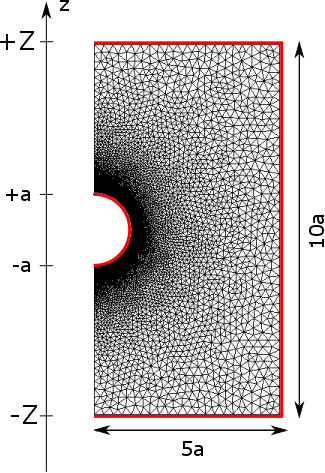}
\caption{Domain of simulation. 
The mesh was generated by \textit{Gmsh} \cite{Gmsh}.
Thick red lines depict physical boundaries of the domain. }\label{Domain}
\end{center}
\end{figure}

\subsection{Alignment tensor}

The non-dimensionalized Landau-de Gennes free energy $\mathcal{E}_{LdG}$, which enters \eqref{The_System_nondim} and subsequently  \eqref{Eq_for_Q} and \eqref{Eq_for_v}, is given by
\begin{equation}\label{E_LdG_nondim}
\mathcal{E}_{LdG} = {\left(\frac{a}{\xi}\right)}^2 \left\{-\frac{1}{2}\text{Tr }\mathsf{Q}^2 +\frac{B}{3A}\text{Tr }\mathsf{Q}^3 +\frac{C}{4A}\left(\text{Tr }\mathsf{Q}^2\right)^2 \right\} +\frac{1}{2}\left|\nabla\mathsf{Q}\right|^2,
\end{equation} 
where $\xi=\sqrt{L/A}\sim 10$~nm stands for the nematic coherence length and $A$, $B$, and $C$ are constant at a given temperature.
The Landau-de Gennes potential $\mathcal{E}_{LdG}^p$ defined in \eqref{E_LdGp} determines whether the nematic phase is thermodynamically stable. It is minimized by a uniaxial tensor $\mathsf{Q}=S_0(\mathbf{n}\otimes\mathbf{n}-\frac{1}{3}\mathsf{I})$ with $S_0=\frac{1}{4C}\left(-B+\sqrt{B^2+24AC} \right)$ for any $\mathbf {n}\in\mathbb S^2$.
Following Fukuda \textit{et al.} \cite{Fukuda_ring, Fukuda_flow}, we set $C=-B=3A$ so as $S_0=1$.
Assuming the same scalar order parameter $S_0=1$ at the particle surface and introducing a unit-length vector $\bm{\nu}$ normal to it, we impose the Dirichlet boundary condition $\mathsf{Q}=\bm{\nu}\otimes\bm{\nu}-\frac{1}{3}\mathsf{I}$ corresponding to the strong homeotropic anchoring of the nematic. At infinity we assume the uniform nematic alignment, i.e., $\mathsf{Q}=\mathbf{n}^0\otimes\mathbf{n}^0-\frac{1}{3}\mathsf{I}$, where $\mathbf{n}^0=(0, 0, 1)$.
The topological constraints imposed by our choice of boundary data produce either a line or point singularity in the vicinity of the particle.
Theoretical \cite{Stark_transition, Fukuda_ring, Ravnik_modelling} and experimental \cite{Loudet,Voltz} studies show that a small particle ($a/\xi\lesssim 60$) will be encircled by a disclination loop, known as a Saturn ring, whereas a point defect, a hyperbolic hedgehog, will be energetically favorable provided that $a/\xi\gtrsim 60$. Note that both configurations are axisymmetric with respect to $\mathbf{n}^0$.
Therefore, in cylindrical coordinates $\{\rho, \phi, z\}$ with the $z$-axis pointing along the director at infinity $\mathbf{n}^0$, the alignment tensor $\mathsf{Q}=\mathsf{Q}(\rho,z)$ does not depend on the azimuthal angle $\phi$. 

While the problem \eqref{Eq_for_Q} was solved explicitly in the limit of small particles \cite{Alama}, there is no analytical solution for the hedgehog configuration in three dimensions. In two dimensions the solution, however, is well known \cite{Lubensky}. Indeed, the director field $\mathbf{n}^{2D}=(\cos\psi, \sin\psi)$ around a circular particle located at the origin of Cartesian coordinate system $\{x,y\}$ and a pointlike topological defect at $(0, -y_0)$ is given by
\begin{equation}\label{Psi}
\psi = 2\arctan\frac{x}{y} -\arctan\frac{x}{y+y_0} -\arctan\frac{x}{y+1/y_0}.
\end{equation} 
In our study, this two-dimensional solution $\mathbf{n}^{2D}$ is used as an initial guess for the axially symmetric problem. We use the nonlinear variational solver developed by the FEniCS Project---a collection of open source software for automated solution of differential equations by finite element methods \cite{AlnaesBlechta2015a, LoggMardalEtAl2012a,LoggWellsEtAl2012a,LoggWells2010a, KirbyLogg2006a,LoggOlgaardEtAl2012a, OlgaardWells2010b, AlnaesEtAl2012, Alnaes2012a, Kirby2004a, Kirby2004a, AlnaesLoggEtAl2009a, AlnaesLoggEtAl2012a}.
In the case of small particles $(a/\xi\lesssim 60)$, the initial state relaxes to a Saturn ring configuration, while for large particles $(a/\xi\gtrsim 60)$ it results in a hedgehog-like solution that, in agreement with \cite{Fukuda_ring, Ravnik_modelling,PhysRevLett.116.147801}, is in fact a small ring disclination rather than a point defect. 

\begin{figure}
\begin{center}
\includegraphics[width=.36\textwidth]{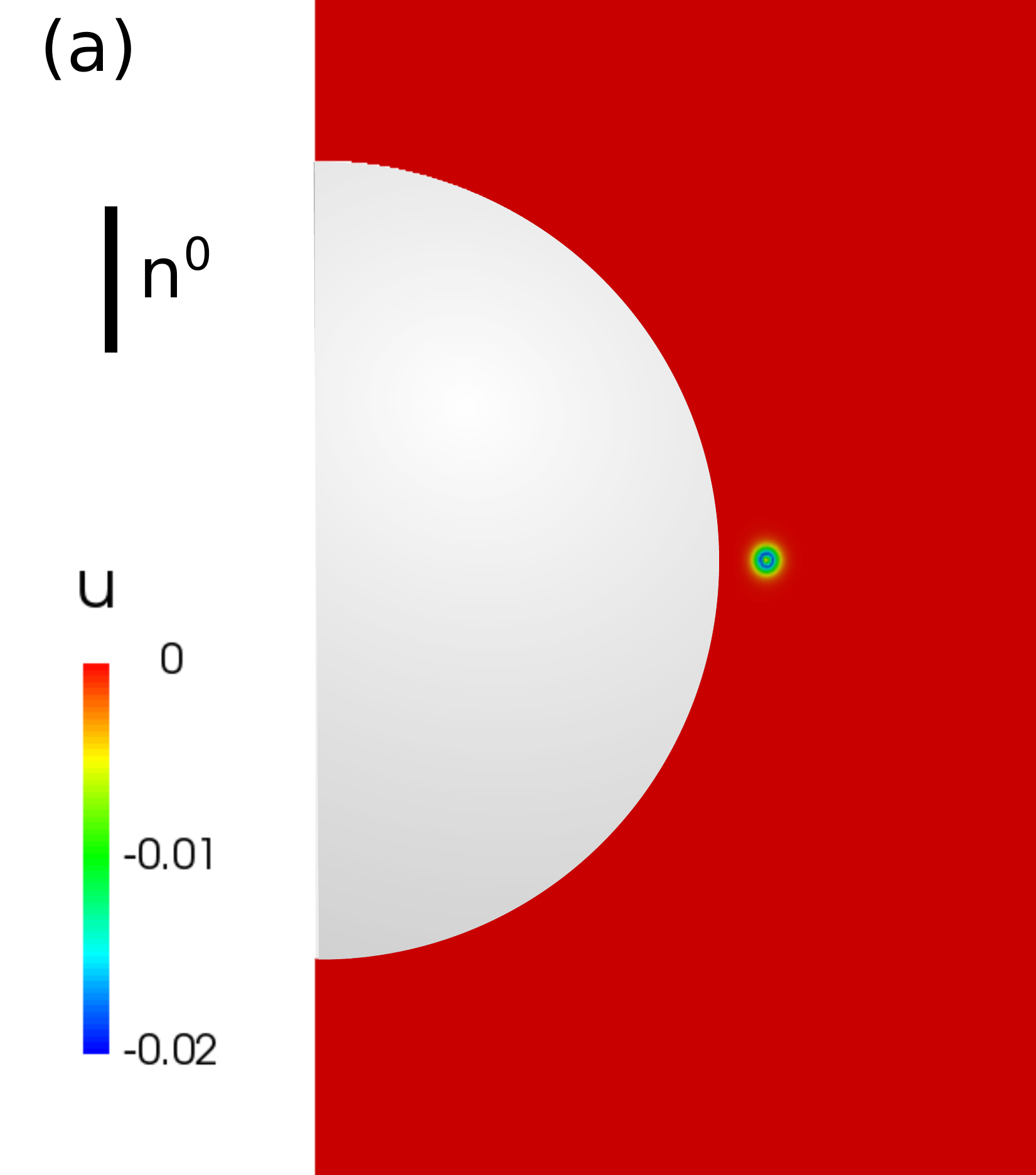}
\hspace{10pt}
\includegraphics[width=.36\textwidth]{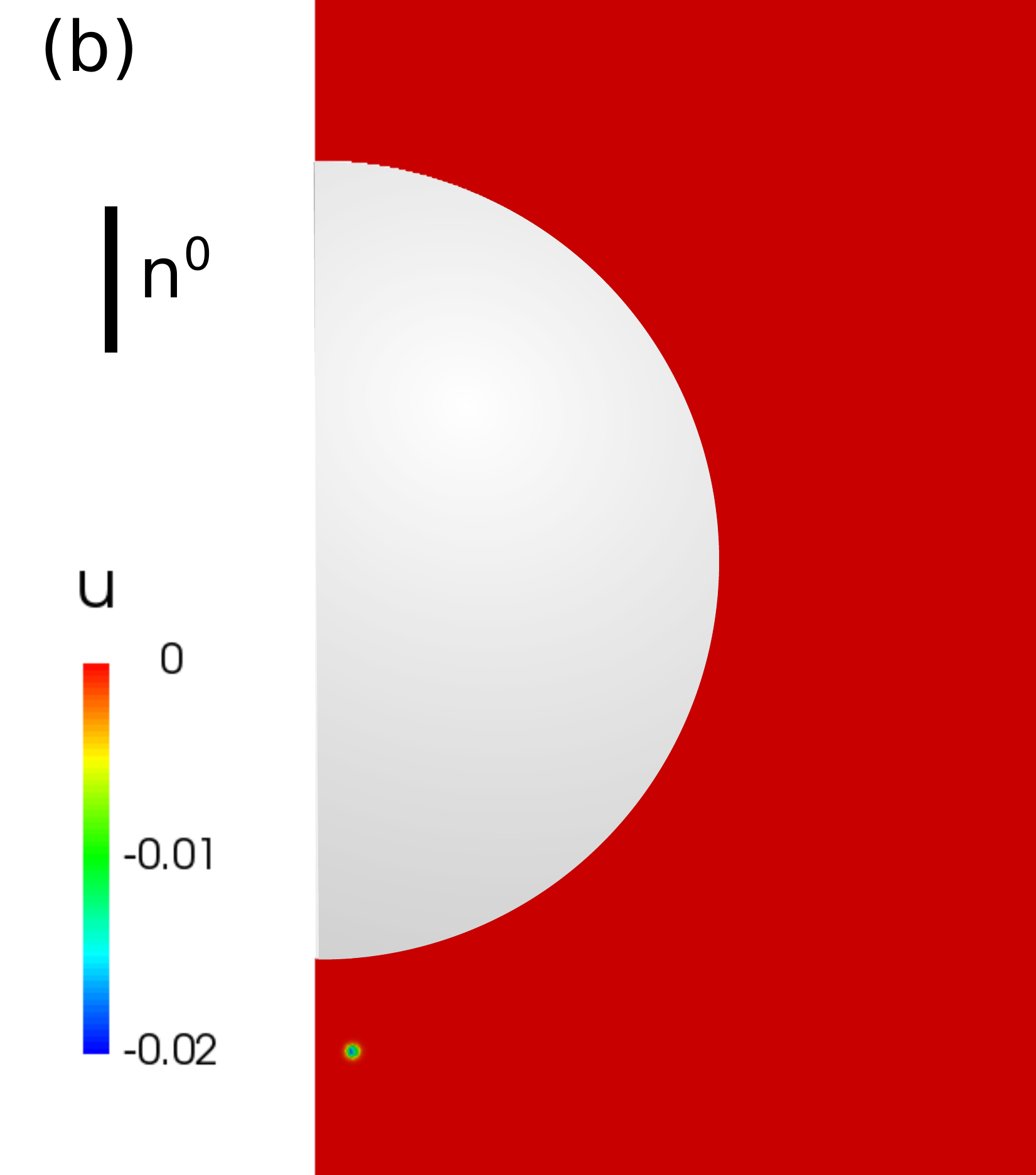}
\caption{Spherical particle accompanied by a Saturn ring \textbf{(a)} and a hyperbolic hedgehog \textbf{(b)} topological defects.
Nonzero values of the biaxiality parameter $u$ given by \eqref{Criterion} indicate biaxial alignment of the liquid crystal molecules.}\label{Criterion_plot}
\end{center}
\end{figure}

The computed solutions of the problem \eqref{Eq_for_Q} for $a/\xi=30$ and $a/\xi=70$ are visualized in Fig.~\ref{Criterion_plot} by plotting of a scalar criterion $u$ proposed in \cite{Lux}. The criterion utilizes the fact that the eigenvalues of the tensor order parameter $\mathsf{Q}$ corresponding to a uniaxial nematic state can be written as $-s$, $-s$, $2s$. Then $\text{Tr }\mathsf{Q}^2=6s^2$ and $\det\mathsf{Q}=2s^3$ and one can introduce a scalar quantity
\begin{equation}\label{Criterion}
u = \frac{\left(\det\mathsf{Q} \right)^2}{\left(\text{Tr }\mathsf{Q}^2 \right)^3} -\frac{1}{54},
\end{equation}
whose nonzero values indicate biaxial alignment of the liquid crystal molecules. Note that in the absolute units, the radius of the colloidal spheres is rather small, 0.3 microns and 0.7 microns, respectively; experiments reported so far deal with bigger spheres, $a=25$ microns \cite{Lazo_2}.

\subsection{Charge separation}

\begin{figure}
\begin{center}
\includegraphics[width=.36\textwidth]{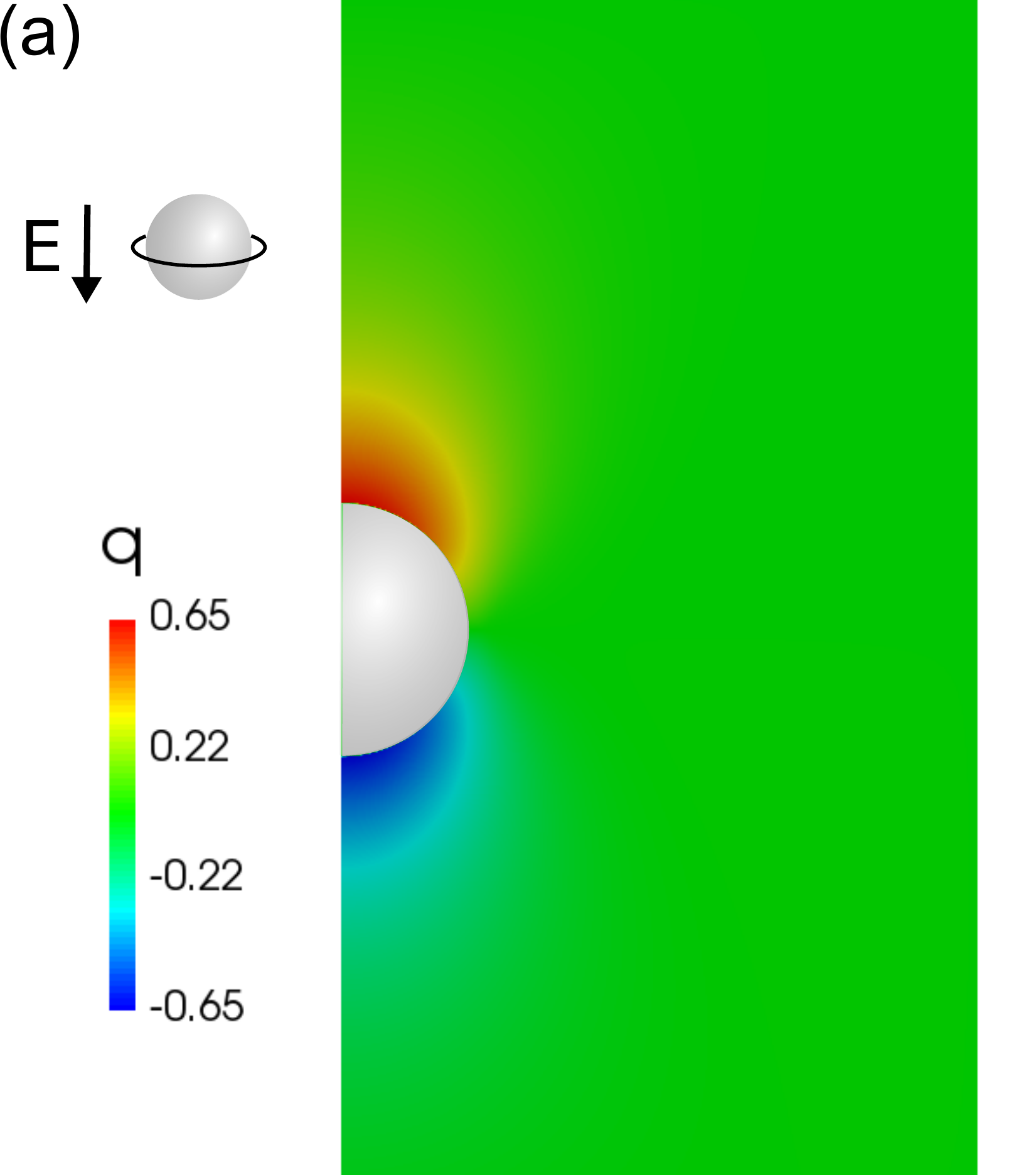}
\includegraphics[width=.36\textwidth]{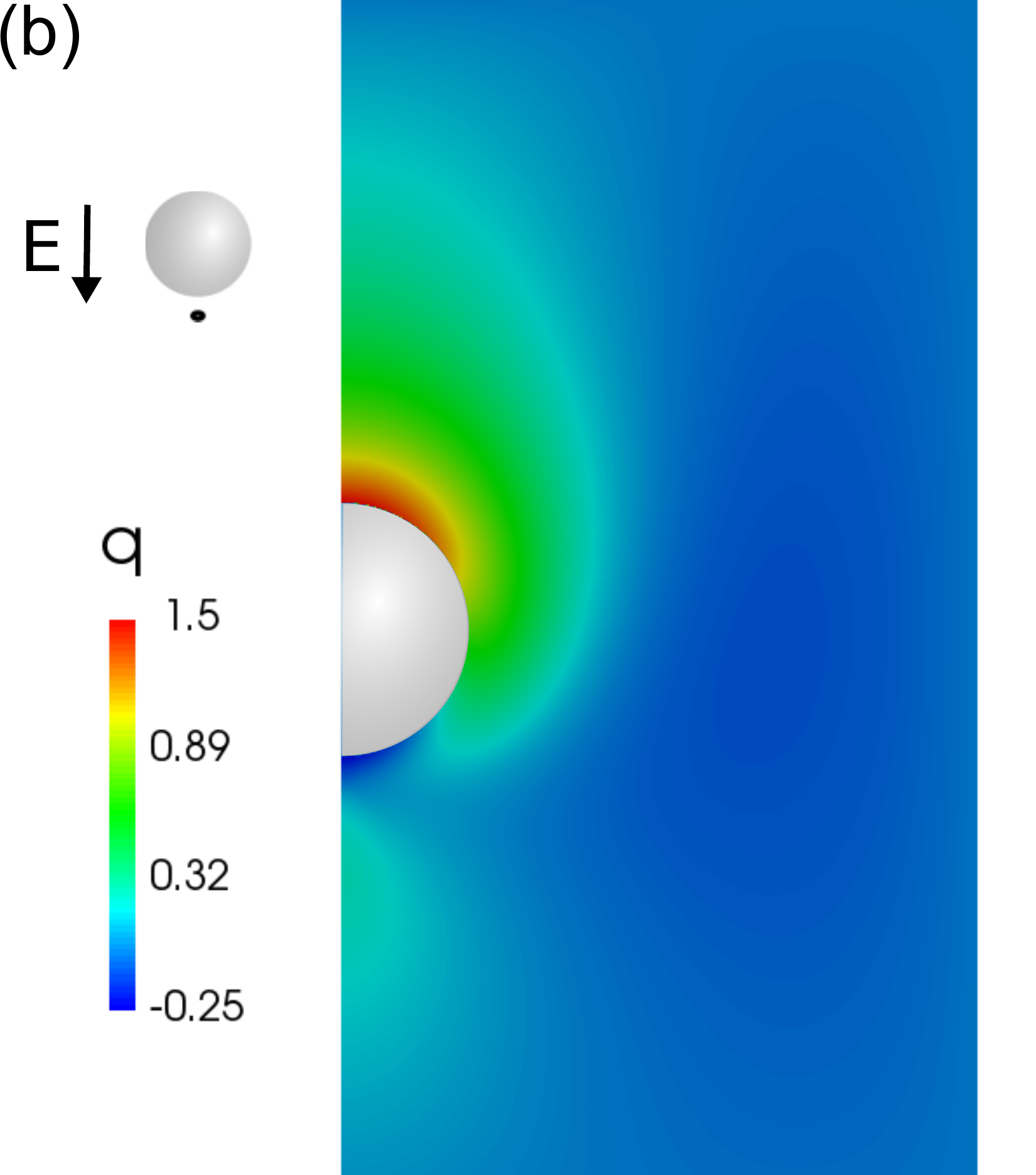}\\
\vspace{3pt}
\includegraphics[width=.36\textwidth]{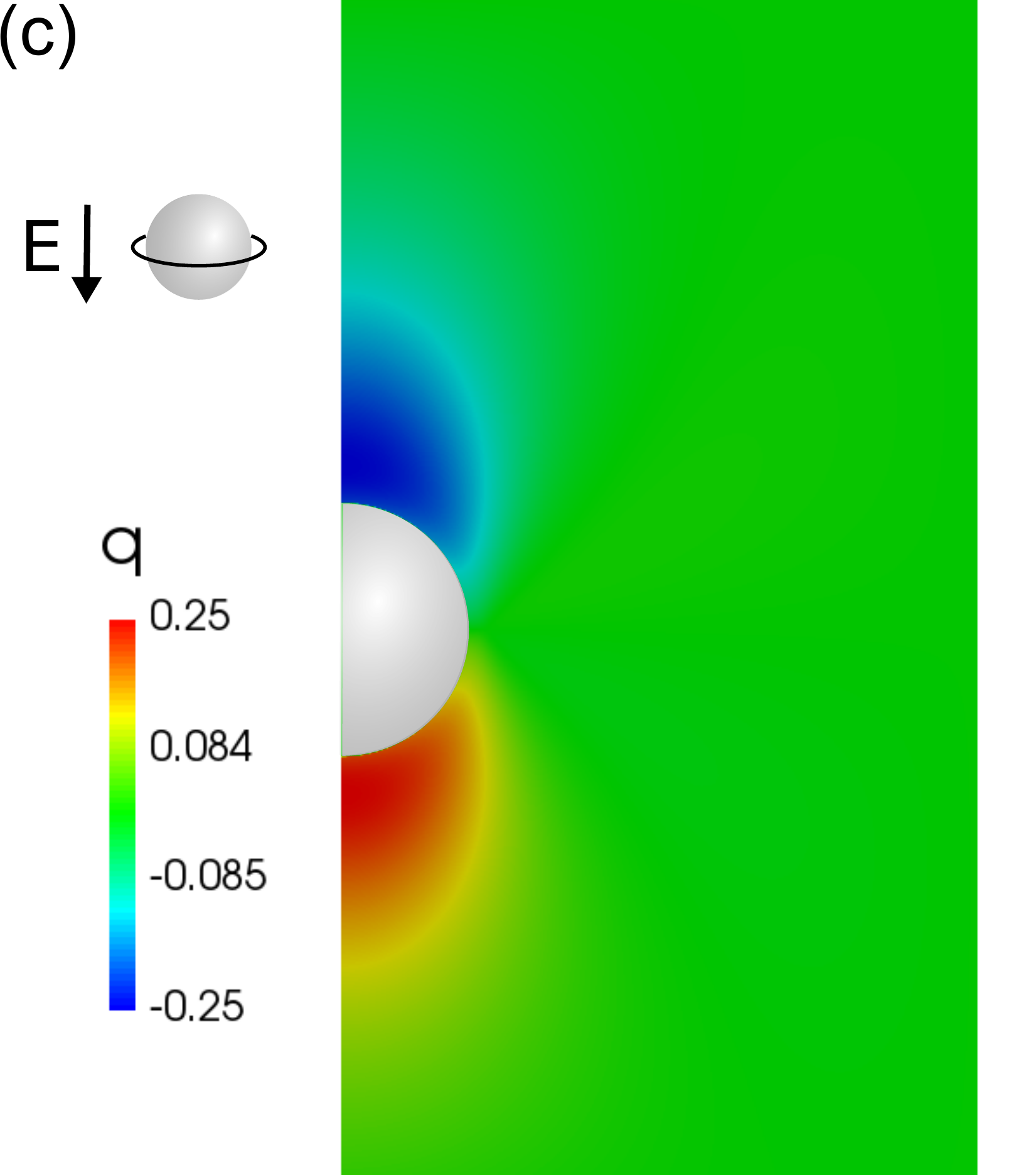}
\includegraphics[width=.36\textwidth]{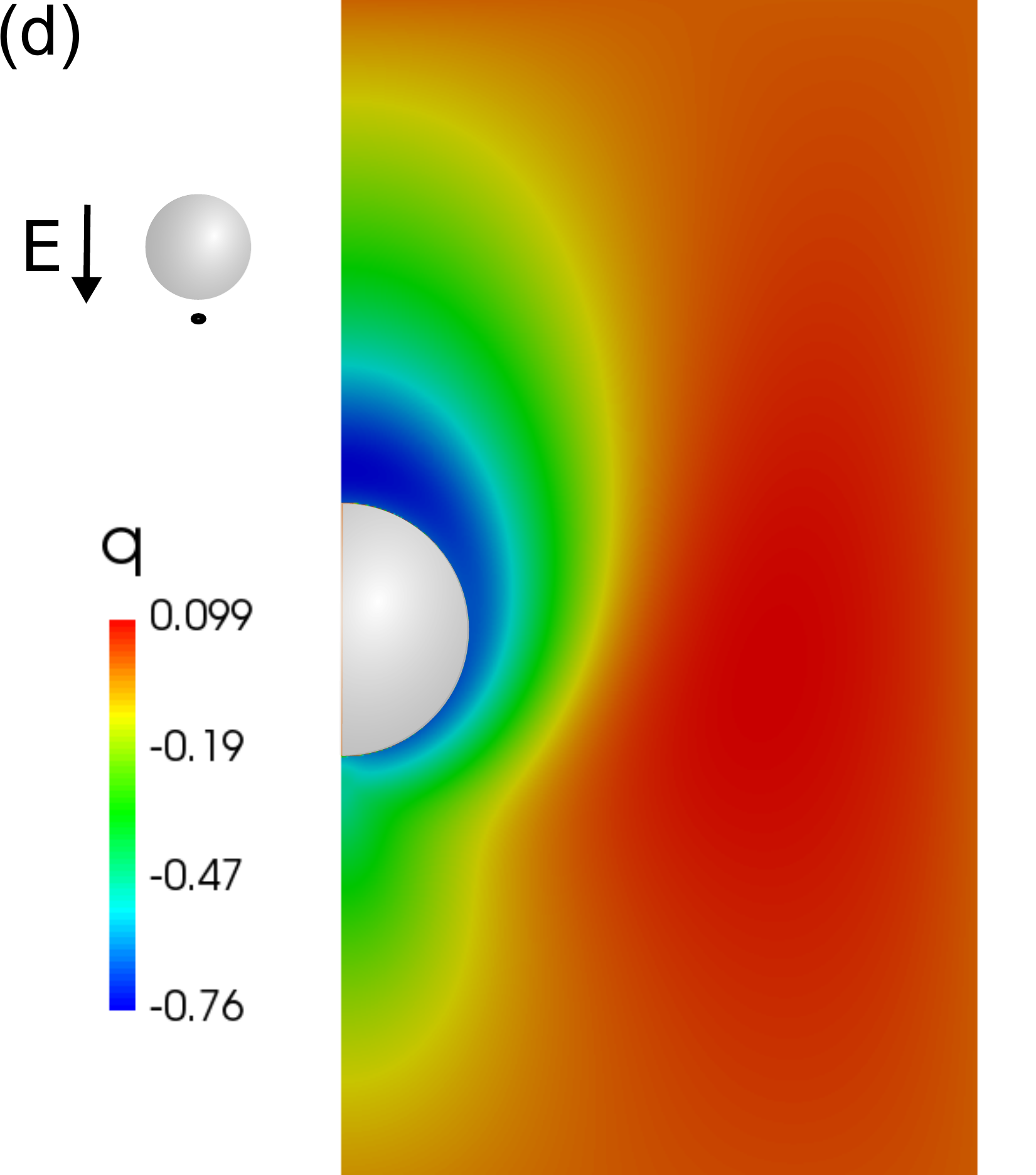}\\
\vspace{3pt}
\includegraphics[width=.36\textwidth]{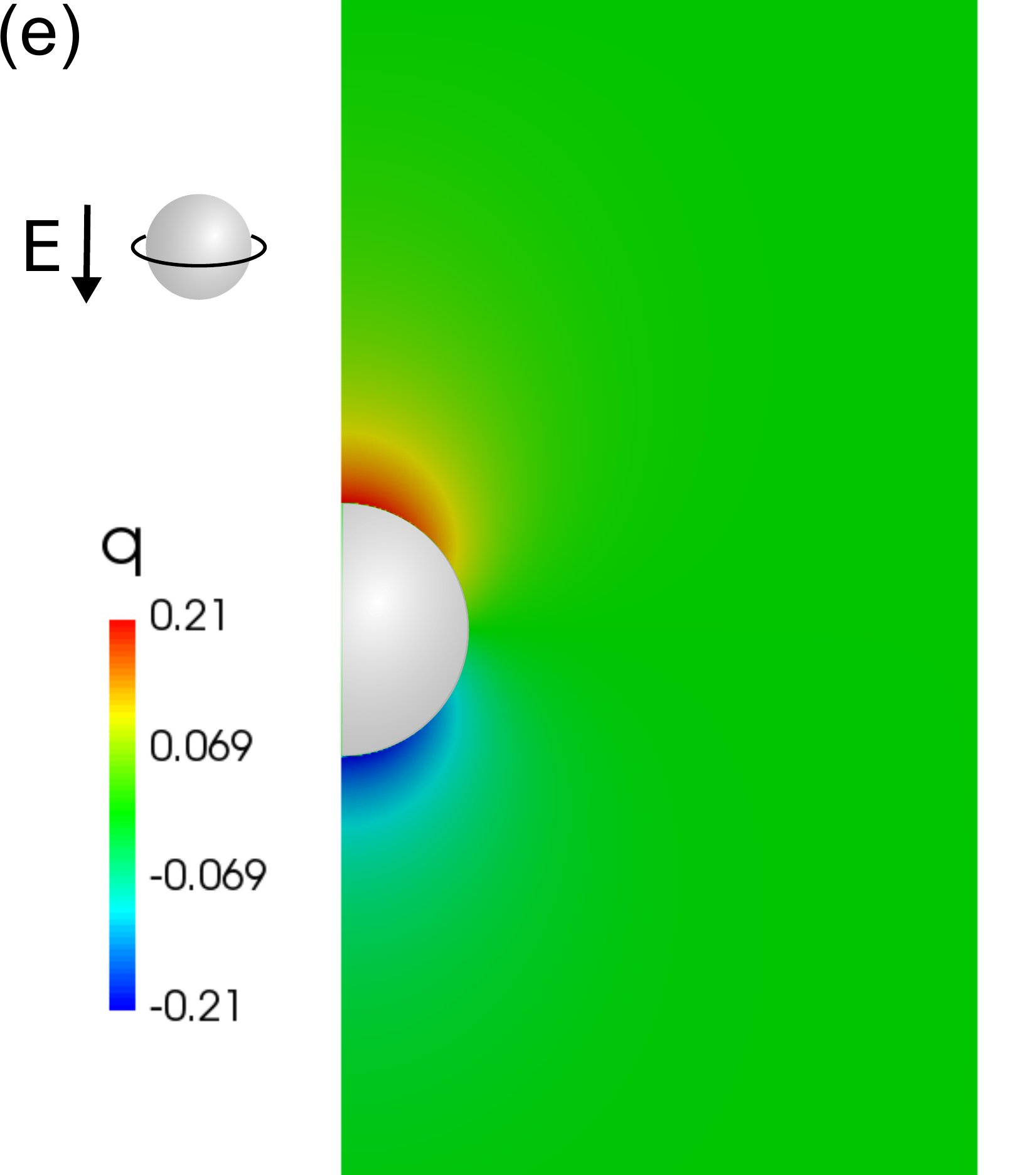}
\includegraphics[width=.36\textwidth]{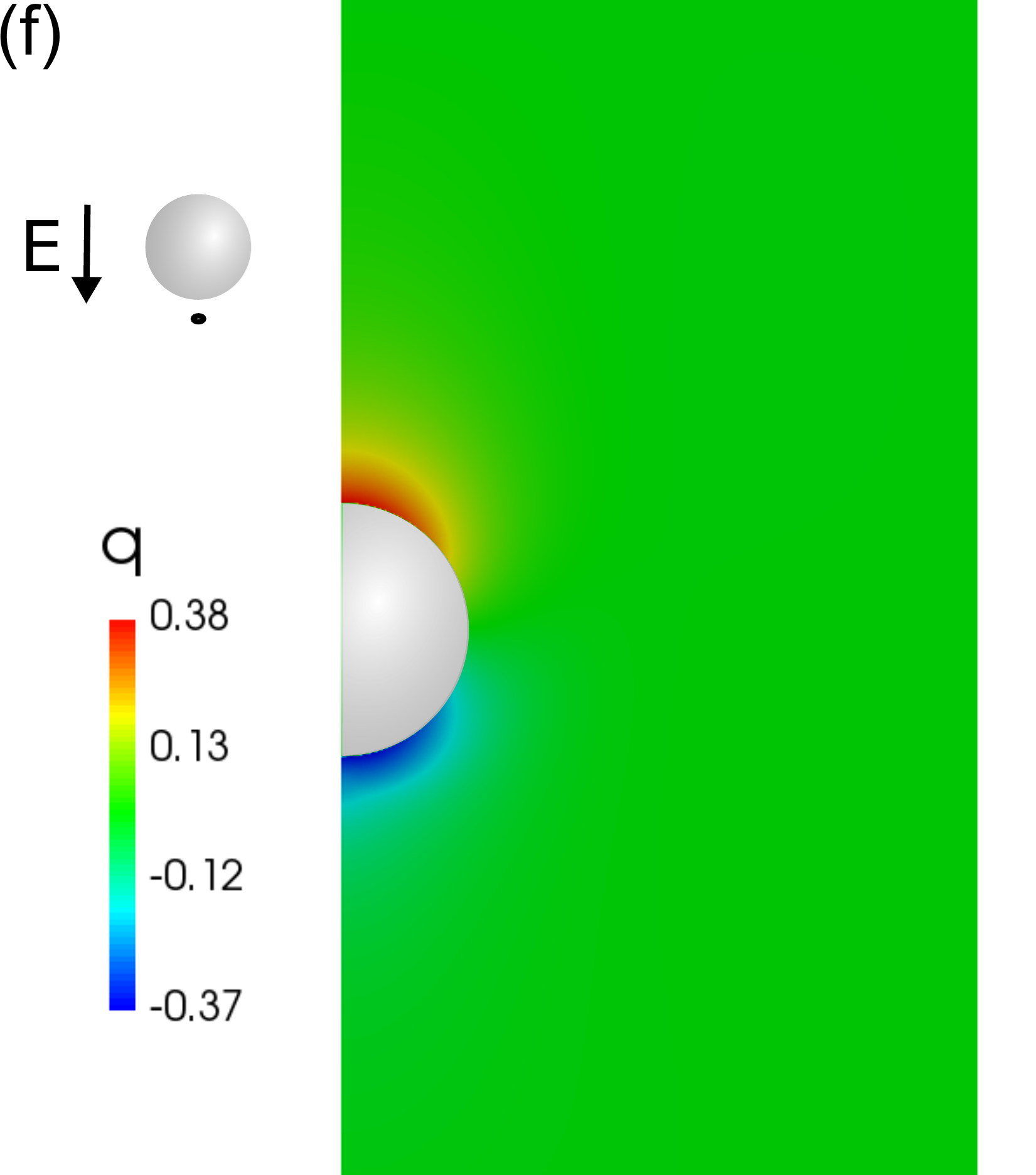}
\caption{Nondimensional charge density $q=c^+-c^-$ around a spherical particle with: a Saturn ring \textbf{(a),(c),(e)}; and a hedgehog \textbf{(b),(d),(f)} topological defect.
Here $\lambda_{\varepsilon}$=1, $\lambda_{\sigma}$=2 in \textbf{(a),(b)};
$\lambda_{\varepsilon}$=2, $\lambda_{\sigma}$=1 in \textbf{(c),(d)}; and
$\lambda_{\varepsilon}=\lambda_{\sigma}$=2 in \textbf{(e),(f)}.}\label{Charge_plot}
\end{center}
\end{figure}

Once the tensor field $\mathsf{Q}$ is known, we solve the problem \eqref{Eq_for_E} for the ionic concentrations $c^{\pm}=c^{\pm}(\rho,z)$ and the electric potential $\Phi=\Phi(\rho,z)$, subject to Dirichlet boundary conditions $c^{\pm}=1$ and $\Phi=z$ at $z=\pm Z$ (see Fig.~\ref{Domain}). Here, the Maxwell equation in \eqref{Eq_for_E}  should also be solved inside the particle. Therefore, the dielectric permittivity $\epsilon_{p}$ of the particle has to be specified as it determines the distribution of ions in the system and thus influences the flow.
In the present study, we focus on dielectric colloids which are commonly used in practice. In particular, Fig.~\ref{Charge_plot} shows nondimensional charge density $q=c^{+}-c^{-}$ around a dielectric spherical particle with $\epsilon_p=0.4\varepsilon_{\perp}$.

Note that the separation of charges in the system arises from an interplay between the orientational ordering of the nematic and its anisotropic permittivity and conductivity, determined by the tensor field $\mathsf{Q}$ and the parameters $\lambda_{\varepsilon}$ and $\lambda_{\sigma}$, respectively. This result is in line with the expectations that the space charge around colloidal spheres is proportional to the anisotropy of dielectric permittivity and electric conductivity \cite{Lazo_2}.  A similar, but probably simpler, interplay in patterned nematics \cite{Carme,Peng_pattern,we_pattern}, where spatially varying director field is induced by means of specific anchoring at the substrates,  yields the electrokinetic charge density $q_{\text{pat}}\propto \lambda_{\varepsilon}-\lambda_{\sigma}$.
In the system under investigation, the charge distribution $q(\mathbf{r})$ is also sensitive to the values of $\lambda_{\sigma}$ and $\lambda_{\varepsilon}$, but it does not vanish when $\lambda_{\varepsilon}=\lambda_{\sigma}$. This is not surprising, given the fact that even in isotropic electrolytes -- where $\lambda_{\varepsilon}=\lambda_{\sigma}=1$ -- a dielectric sphere in presence of an applied electric field is capable of generating space charges and cause induced-charge electro-osmosis (ICEO) \cite{Bazant,gamayunov1986pair,murtsovkin1996nonlinear}.  This effect is especially pronounced when the Debye screening length $\lambda_D=\frac{1}{e}\sqrt{\frac{\varepsilon_0\varepsilon_{medium}k_B\theta}{n}}$ (where $n$ is the concentration of ions) around the colloid is comparable to the radius of the colloid, as will be discussed later in the context of the field-induced electro-osmotic velocities. 

\begin{figure}
\begin{center}
\includegraphics[width=.36\textwidth]{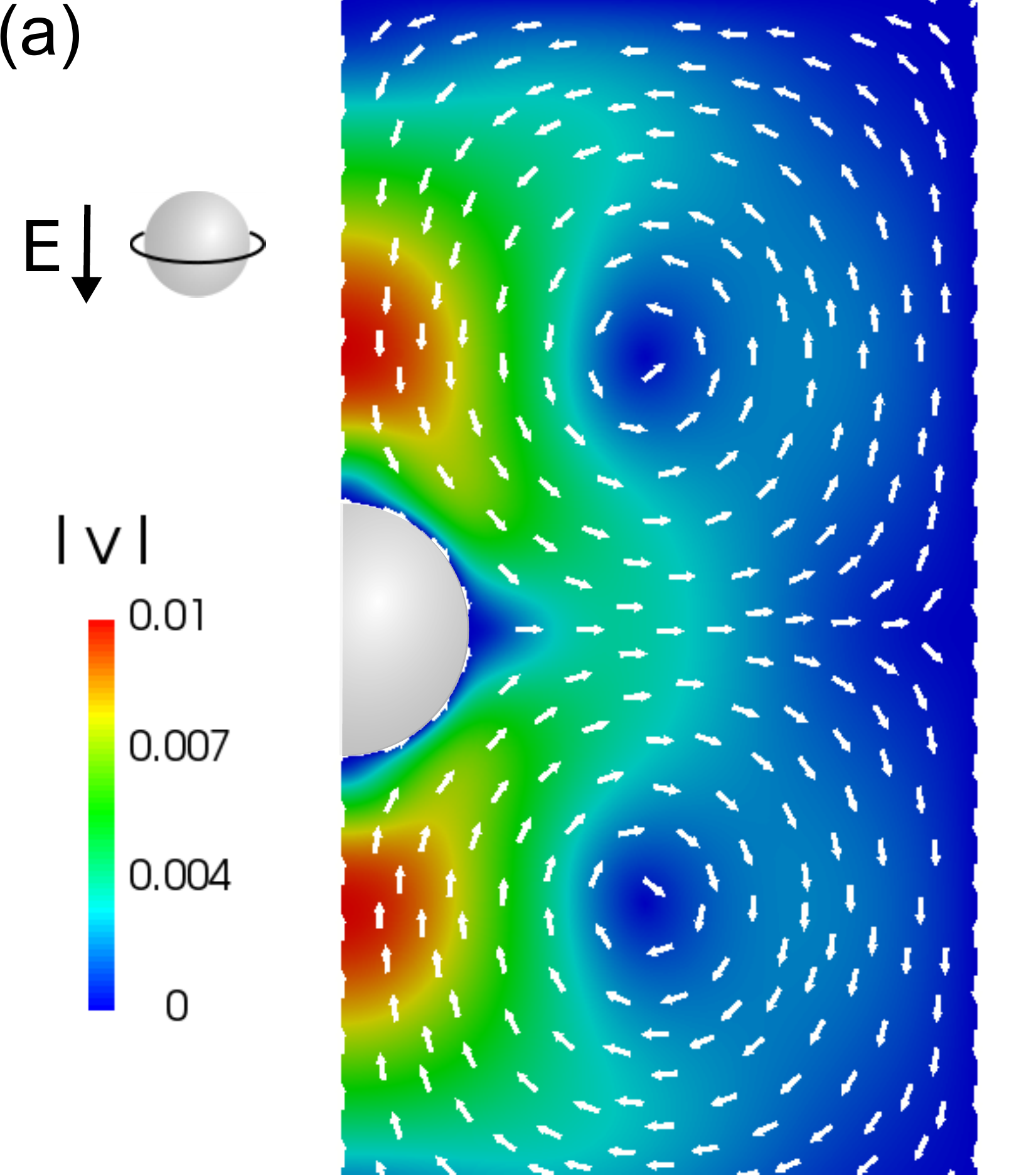}
\includegraphics[width=.36\textwidth]{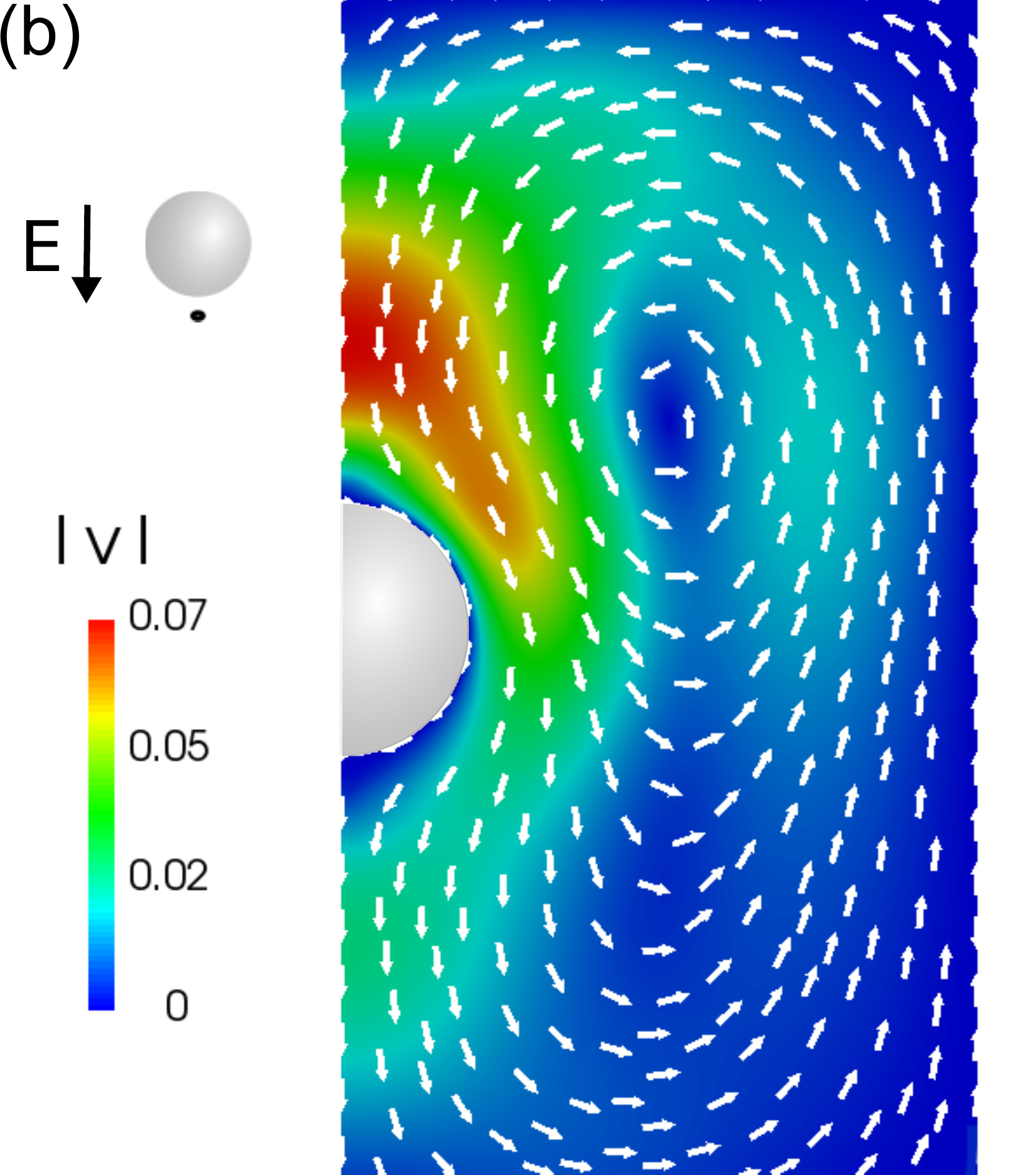}\\
\vspace{3pt}
\includegraphics[width=.36\textwidth]{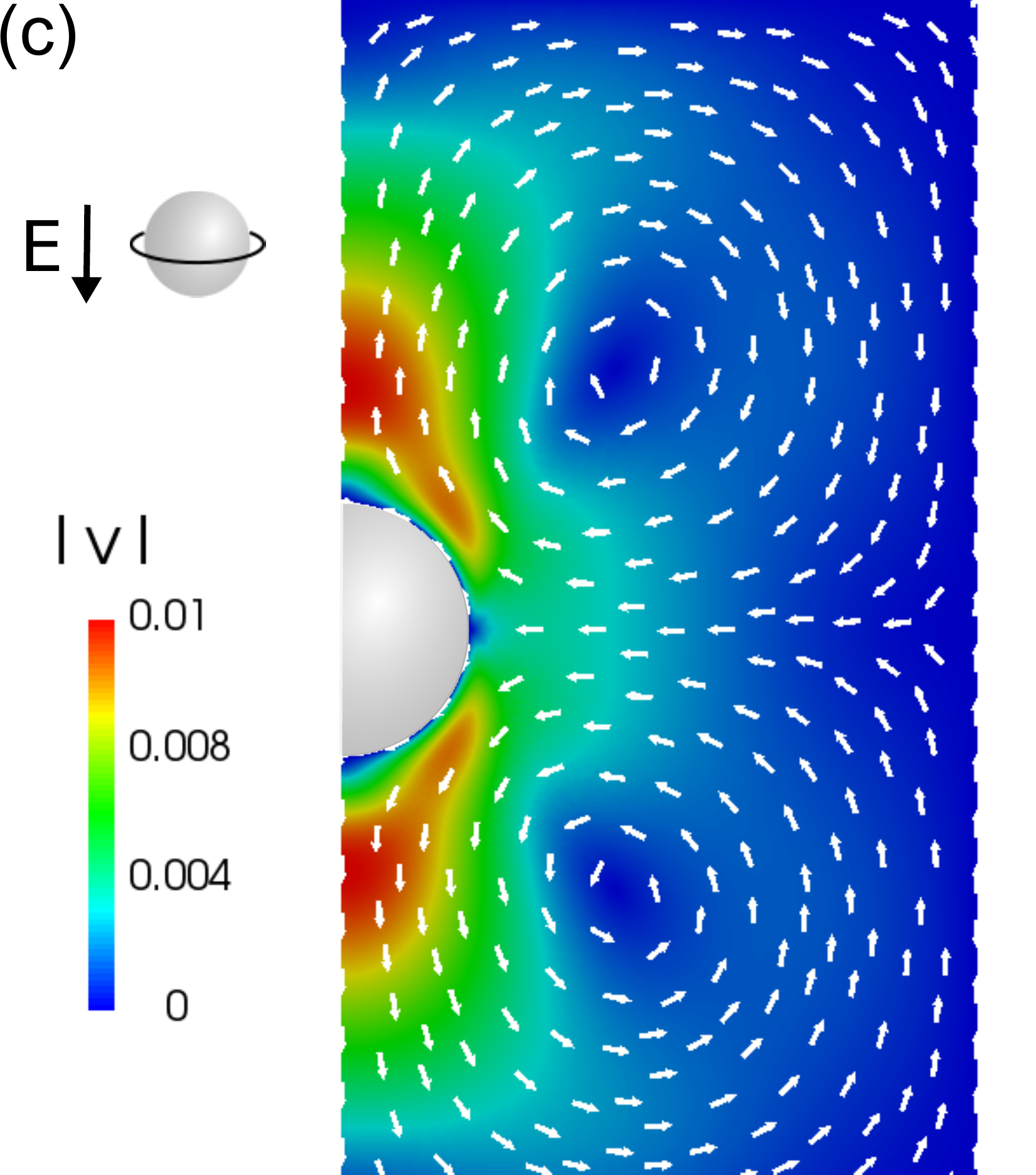}
\includegraphics[width=.36\textwidth]{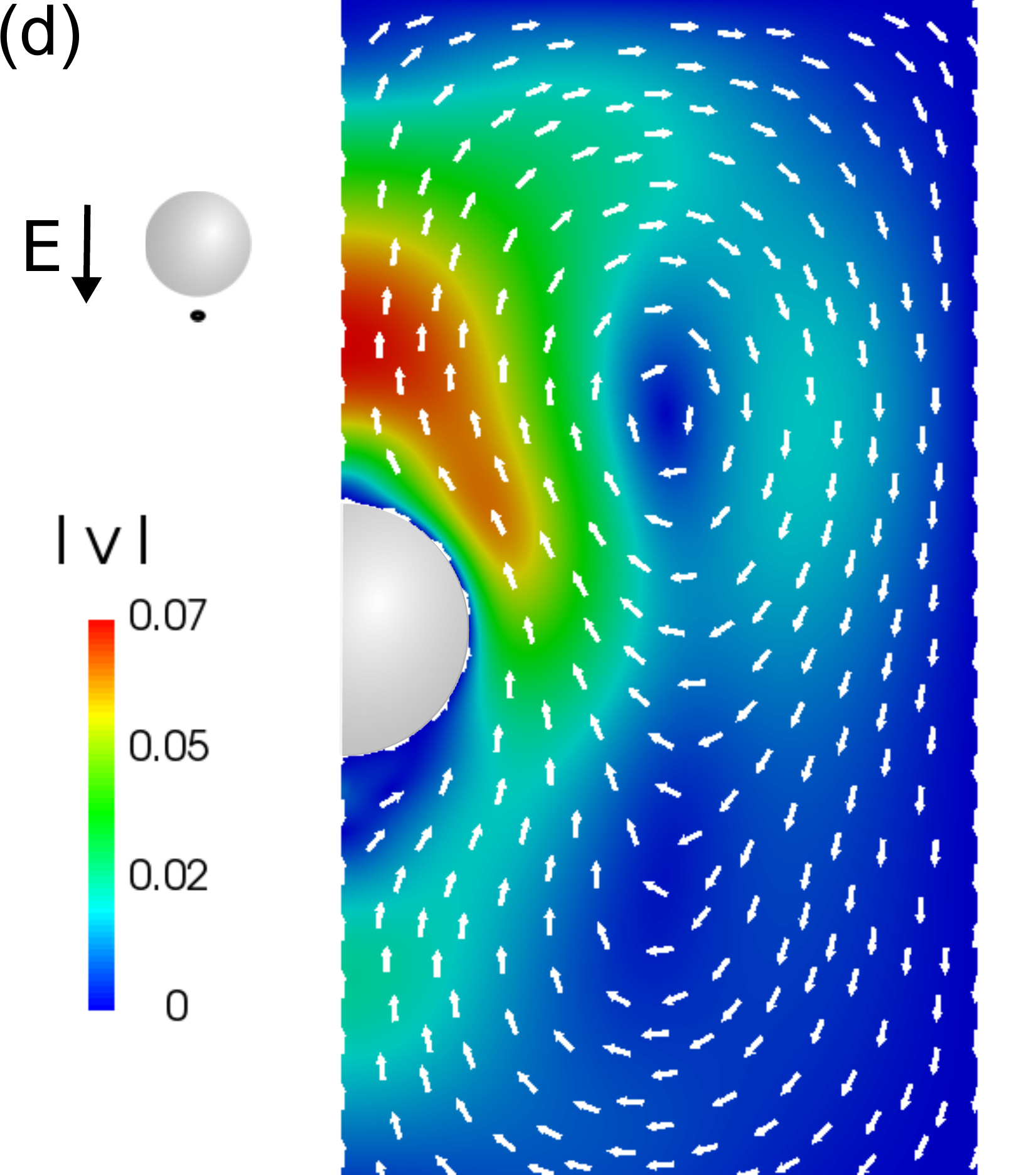}\\
\vspace{3pt}
\includegraphics[width=.36\textwidth]{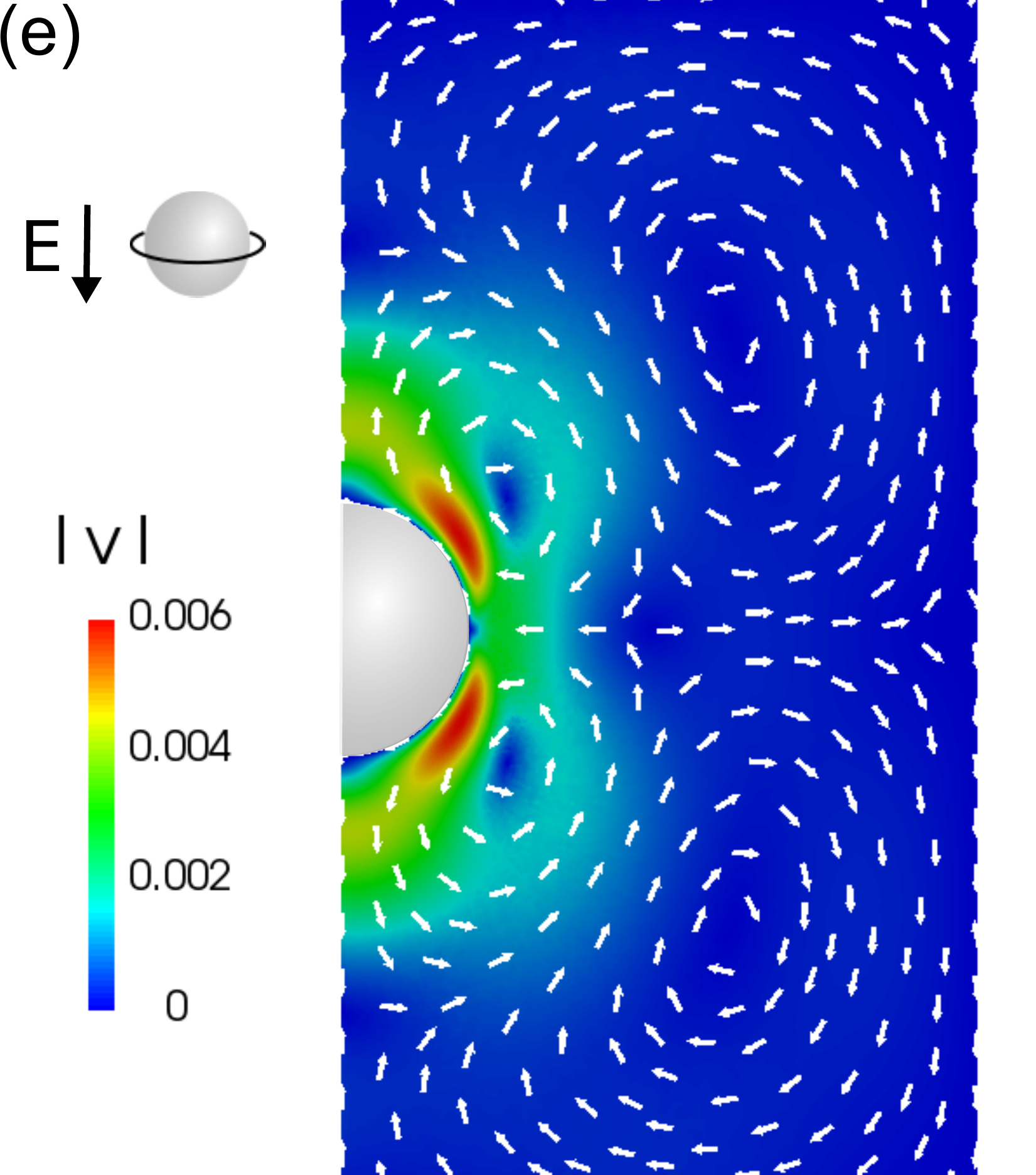}
\includegraphics[width=.36\textwidth]{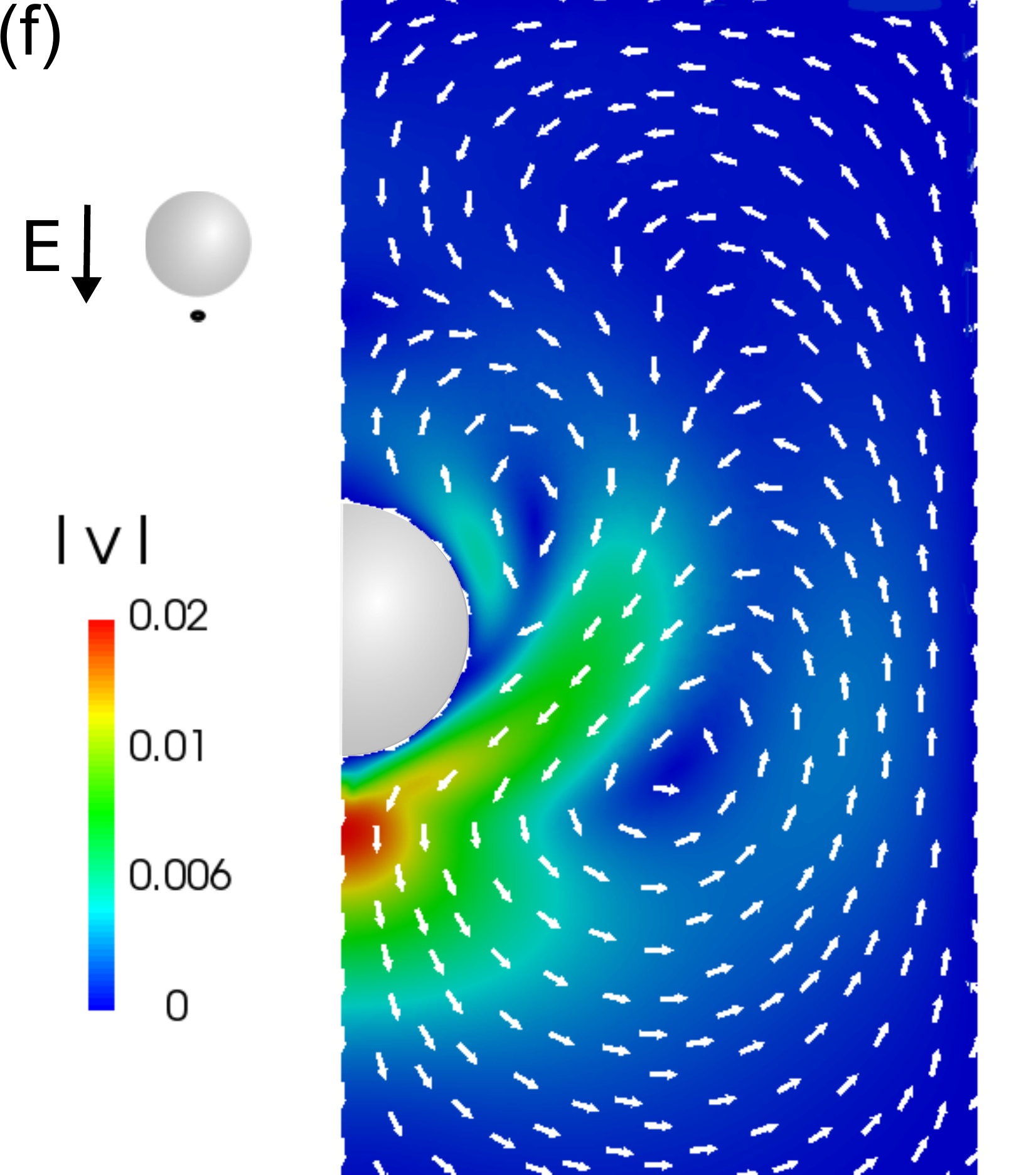}
\caption{Velocity field around a spherical particle with: a Saturn ring \textbf{(a),(c),(e)}; and a hedgehog \textbf{(b),(d),(f)} defect.
Here $\lambda_{\varepsilon}$=1, $\lambda_{\sigma}$=2 in \textbf{(a),(b)};
$\lambda_{\varepsilon}$=2, $\lambda_{\sigma}$=1 in \textbf{(c),(d)}; and
$\lambda_{\varepsilon}=\lambda_{\sigma}$=2 in \textbf{(e),(f)}.
The nondimensional viscosities are as follows: $\tilde{\zeta}_1=0.3$, $\tilde{\zeta}_2=0$, $\tilde{\zeta}_4=1.3$, $\tilde{\zeta}_6=-0.15$.
}\label{Velocity_plot}
\end{center}
\end{figure}

\subsection{Flow profile}     

We are now in a position to solve the system of equations \eqref{Eq_for_v} for the pressure $p=p(\rho,z)$ and the velocity $\mathbf{v}=\mathbf{v}(\rho,z)$ of the electro-osmotic flow.
One can further simplify the problem by taking advantage
of the fact that $\text{Er} \ll 1$ and $a^2 / \xi_E^2 \ll 1$.
Since these two parameters are small, the elastic stress tensor $\mathsf{T}^{\text{el}}=-\dfrac{\partial \mathcal{E}_{LdG}}{\partial(\partial_k \mathsf{Q}_{mn})}(\partial_i \mathsf{Q}_{mn})\,\mathbf{e}_i\otimes\mathbf{e}_k$ is determined by the order parameter $\mathsf{Q}$ that satisfies \eqref{Eq_for_Q}.
It follows then that $\text{div} \mathsf{T}^{\text{el}}=-\nabla\mathcal{E}_{LdG}$.
Now splitting the total pressure $p$ into the static $p^0=\text{const}-\mathcal{E}_{LdG}/\text{Er}$ and hydrodynamic $p^h$ parts \cite{Stark_Stokes}, we arrive at the following system 
\begin{equation}\label{Eq_for_flow}
\begin{cases}
\text{div} \left[ p^h\mathsf{I} -\mathsf{T}^V -\frac{1}{\varepsilon_{\perp}} \mathbf{E}\otimes\hat{\varepsilon}\mathbf{E} \right] =0,\\
\text{div } \mathbf{v} = 0.
\end{cases}
\end{equation}
Here the viscous stress is
\begin{equation}
\mathsf{T}^V = \tilde{\zeta}_1 \left(\mathsf{Q}\mathring{\mathsf{Q}}-\mathring{\mathsf{Q}}\mathsf{Q}\right) +\tilde{\zeta}_2\mathring{\mathsf{Q}} +\left(\tilde{\zeta}_4+\tilde{\zeta}_2\right)\mathsf{Q}\mathsf{A} +\left(\tilde{\zeta}_4-\tilde{\zeta}_2\right)\mathsf{A}\mathsf{Q} +\tilde{\zeta}_6 \text{Tr} \left(\mathsf{Q}\mathsf{A}\right)\mathsf{Q} +\mathsf{A},
\end{equation}
where $\tilde{\zeta}=\zeta/\zeta_8$, $2\mathsf{A}=\nabla\mathbf{v}+(\nabla\mathbf{v})^T$, and $\mathring{\mathsf{Q}} = \dot{\mathsf{Q}} -\mathsf{WQ} +\mathsf{QW}$ with $2\mathsf{W}=\nabla\mathbf{v}-(\nabla\mathbf{v})^T$.

Solutions to \eqref{Eq_for_flow}  computed under no-slip conditions $(\mathbf{v}=0)$ at the physical boundaries of the domain of simulation (see Fig.~\ref{Domain})  are depicted in Fig.~\ref{Velocity_plot}. 
Similar to the charge density $q$ discussed above, the flow $\mathbf{v}$ is sensitive to the degrees of anisotropy $\lambda_{\varepsilon}$ and $\lambda_{\sigma}$, as well as to the symmetry of the director field.
In particular, the { quadrupolar} flow profiles around the particle encircled by an equatorial Saturn ring are symmetric with respect to the plane of the defect. 
On the contrary, the particle accompanied by a hedgehog gives rise to the velocity fields $\mathbf{v}$ of dipolar symmetry, in qualitative agreement with \cite{Lazo_2}. {Indeed, the direct comparison can be made between the Fourier analysis of the experimental velocity data in Fig.~4 in \cite{Lazo_2} and the insets (a) and (b) in Fig.~\ref{Velocity_plot}, given that $\lambda_\sigma>1$ and $\lambda_\varepsilon=1$ in both cases. The flow profiles in Fig.~4c around the sphere with a disclination ring, shown in \cite{Lazo_2} and Fig.~\ref{Velocity_plot}a are both of the "puller" type with the streams along the axis parallel to the electric field being directed toward the sphere. The flow in Fig.~\ref{Velocity_plot}a consists of the two rolls, that are also present in Fig.~4c in \cite{Lazo_2}. The experiment also shows pairs of micro-vortices located very closely to the poles of the sphere of a size that is smaller than the radius of the sphere. These microvortices are not featured in the simulations, apparently because of the differences between the confinement geometries considered here and in \cite{Lazo_2}. Note that the quadrupolar symmetry of the director pattern in the disclination ring configuration makes the electro-osmotic flows symmetric with respect to the  equatorial plane of the sphere.  There is thus no "pumping" of the fluid from one pole of the sphere to another, as demonstrated experimentally in \cite{Lazo_2}. The situation changes for the sphere with an accompanying hedgehog, as described below.

The flow profiles around the sphere with a dipolar director configuration caused by the hedgehog are of the "pumping" type in both the experiments (Fig.~4f in \cite{Lazo_2}) and simulations (Fig.~\ref{Velocity_plot}a), with the mirror symmetry with respect to the equatorial plane being broken. The flow in Fig.~\ref{Velocity_plot}b consists of one roll. The flow at the axis of rotational symmetry of the configuration is directed from the side that is defect free to the surface of the sphere. The maximum velocity of the axial flow is achieved at the defect-free side of the sphere; the axial velocity is much lower near the hedgehog. All these features are in complete agreement with the experiment, see Fig.~4f in \cite{Lazo_2}. The vortex in Fig.~\ref{Velocity_plot}b rotates in the counterclockwise direction; its center is shifted towards the defect-free end of the sphere, again as in the experiment \cite{Lazo_2}. The only difference is that the experiment shows an additional vortex in a far field, with the center that is separated from the sphere by a distance about $4a$; this vortex does not appear in the simulations, apparently because of the difference in the confinement geometry (note that in addition to being shallow, the experimental cell is practically infinitely long and wide in the horizontal plane, which brings another difference as compared to the domain of simulations). 

Interchanging the values of $\lambda_\sigma$ and $\lambda_\varepsilon$ in Fig.~\ref{Velocity_plot}c,d essentially reverses the direction of the flow, confirming the observation that the velocity in LCEO should be proportional to the difference between these quantities at leading order \cite{Lazo_2}. This reversal is also in agreement with the recent experiments and 2D director-based numerical simulations \cite{Paladugu} performed for a liquid crystal in which the sign of $\lambda_\sigma-\lambda_\varepsilon$ can be reversed by a suitable choice of composition or temperature. However, if one extends the comparison of the present simulations to the experimental LCEO flows in patterned nematic cells without colloidal inclusions \cite{Carme,Peng_pattern, we_pattern}, then one can observe an important difference.  Namely, the LCEO flows in patterned nematics \cite{Carme,Peng_pattern, we_pattern} vanish when $\lambda_\varepsilon$ and $\lambda_\sigma$ are equal. In contrast, our simulations demonstrate nonzero velocity field $\mathbf{v}$ even in the case of $\lambda_{\varepsilon}=\lambda_{\sigma}$. As mentioned above, this effect is in line with the model developed for ICEO flows around dielectric spheres \cite{Bazant,gamayunov1986pair,murtsovkin1996nonlinear}.  We now discuss this issue in a greater detail.}

Considering an uncharged immobilized dielectric sphere placed in a uniform electric field, Murtsovkin found the analytical solutions for the radial and azimuthal ICEO flows that show a quadrupolar symmetry \cite{murtsovkin1996nonlinear} and a typical amplitude near the surface
\begin{equation}
\label{eq:add}
v^{diel}=\beta\frac{\varepsilon_0\varepsilon_{medium}}{\eta}\frac{aE^2}{1+\frac{\varepsilon_{medium}a}{\varepsilon_p\lambda_D}},
\end{equation}
where $\beta$ is a scalar coefficient that depends on the geometry of the system (for an infinite system with $\lambda_D\ll a$ and $\beta=\frac{9}{32\pi}\approx0.1$).  For an aqueous electrolyte we have that $\varepsilon_{medium}\approx80$, $\lambda_D\approx50$ nm, thus for a typical dielectric particle of a micron size and a permittivity of glass,  $\varepsilon_p\approx5$ , one can safely assume $\varepsilon_{medium}a\gg\varepsilon_p\lambda_D$ so that $v^{diel}=\beta\frac{\varepsilon_0\varepsilon_p}{\eta}\lambda_DE^2$. This velocity is, by a factor about $\lambda_D/a$, smaller than the ICEO flow velocities around ideally polarizable (conductive) spheres \cite{Bazant,murtsovkin1996nonlinear}.  The smallness of this effect around dielectric spheres has been  confirmed experimentally by a  direct comparison of ICEO velocities around conducting (gold) and dielectric (glass) spheres of the same size in the same aqueous electrolyte \cite{peng2014induced}.  In the case of a nematic electrolyte, the ratio $\varepsilon_{medium}a/\varepsilon_p\lambda_D$ is not necessarily very large, as $\varepsilon_{medium}$ and $\varepsilon_p$ are often of the same order of magnitude and the Debye screening length is in the range $0.1-1\,\upmu$m  \cite{thurston1984physical,nazarenko1994anchoring,ciuchi2007ac}. For the micron-size particles considered in this study, $\varepsilon_{medium}a/\varepsilon_p\lambda_D$ is of order $1$.  On the other hand, analytical estimates of the {LCEO} flows velocities yield a typical amplitude $v^{LCEO}=\alpha\frac{\varepsilon_0\varepsilon_\perp}{\eta}\left(\frac{\varepsilon}{\varepsilon_\perp}-\frac{\sigma}{\sigma_\perp}\right)aE^2$ where $\alpha$ is an unknown dimensionless parameter of order $0.1-1$ that is expected to depend on the director field, strength of anchoring, etc. \cite{Lazo_2}. Recent experiments \cite{Paladugu} on LCEP of spheres with $a=5\upmu$m show that $\alpha$ approximately equals $1$. The ICEO and LCEO flow velocities around dielectric spheres in the nematic electrolyte can thus be of comparable magnitudes. When $\frac{\varepsilon}{\varepsilon_\perp}-\frac{\sigma}{\sigma_\perp}=0$, the total velocity around the sphere would not vanish, being determined by the isotropic contribution \eqref{eq:add}. For example, with $\varepsilon_{medium}=\varepsilon_p=7$, $\eta=0.1$ Pa\,s, $a=\lambda_D=0.3\,\upmu$m, $E=40\times10^3\,$V/m, the estimate is $v^{diel}=0.01\,\upmu$m/s.  The ICEO effect is apparently more pronounced around smaller particles explored in this work; as the particles become larger as in the experiments \cite{Lazo_2}, this effect would become of a lesser importance.  On the other hand, the {LCEO} effect is expected to diminish as the particle becomes smaller, since the smaller (submicrometer and less) particles are not capable to produce strong director gradients needed for charge separation.  It would be of interest to explore the relative strength of ICEK and LCEK in the isotropic and the nematic phases of the same liquid crystal material for particles of a different size. 
 
\begin{figure}
\begin{center}
\includegraphics[width=.36\textwidth]{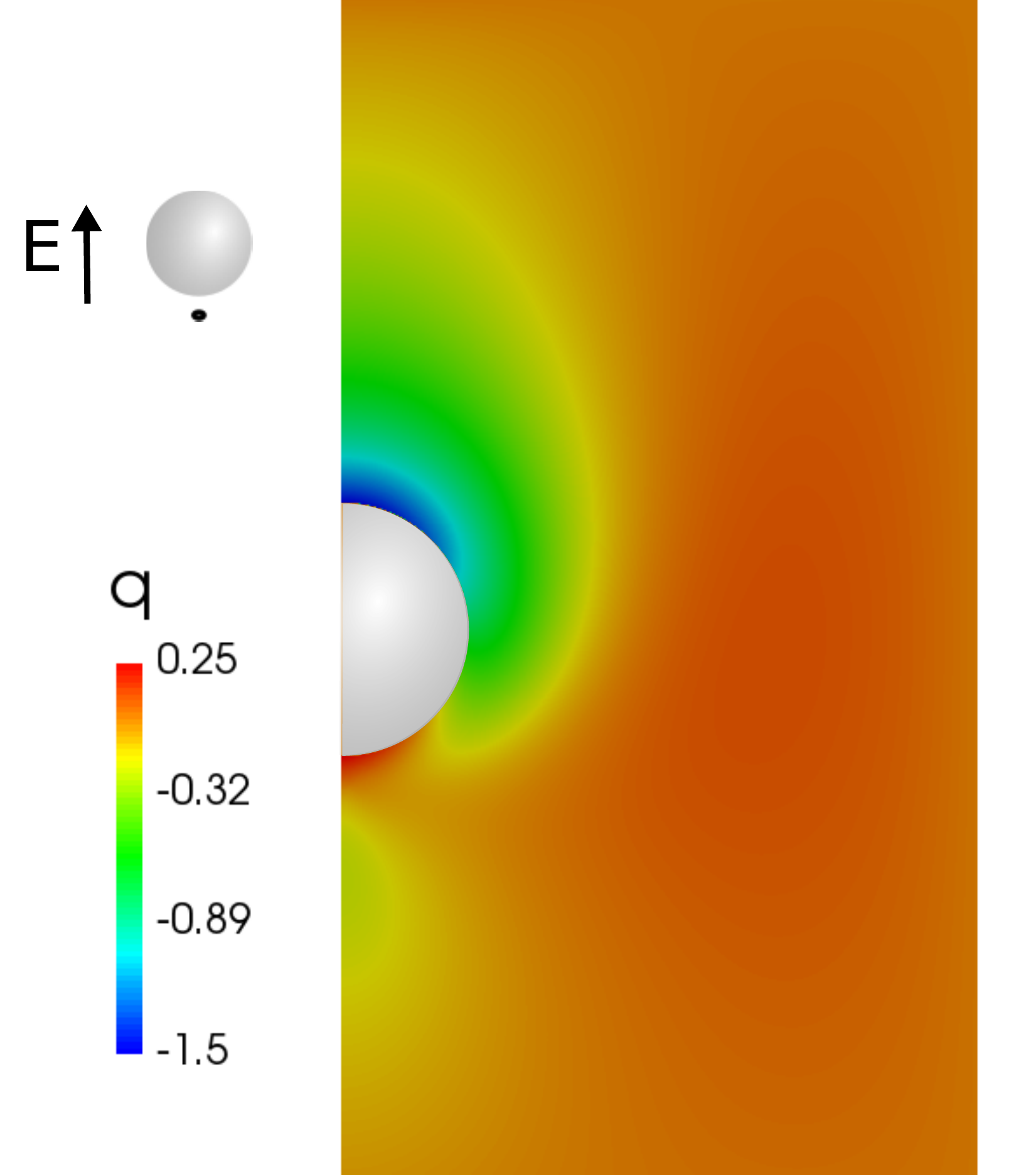}
\includegraphics[width=.36\textwidth]{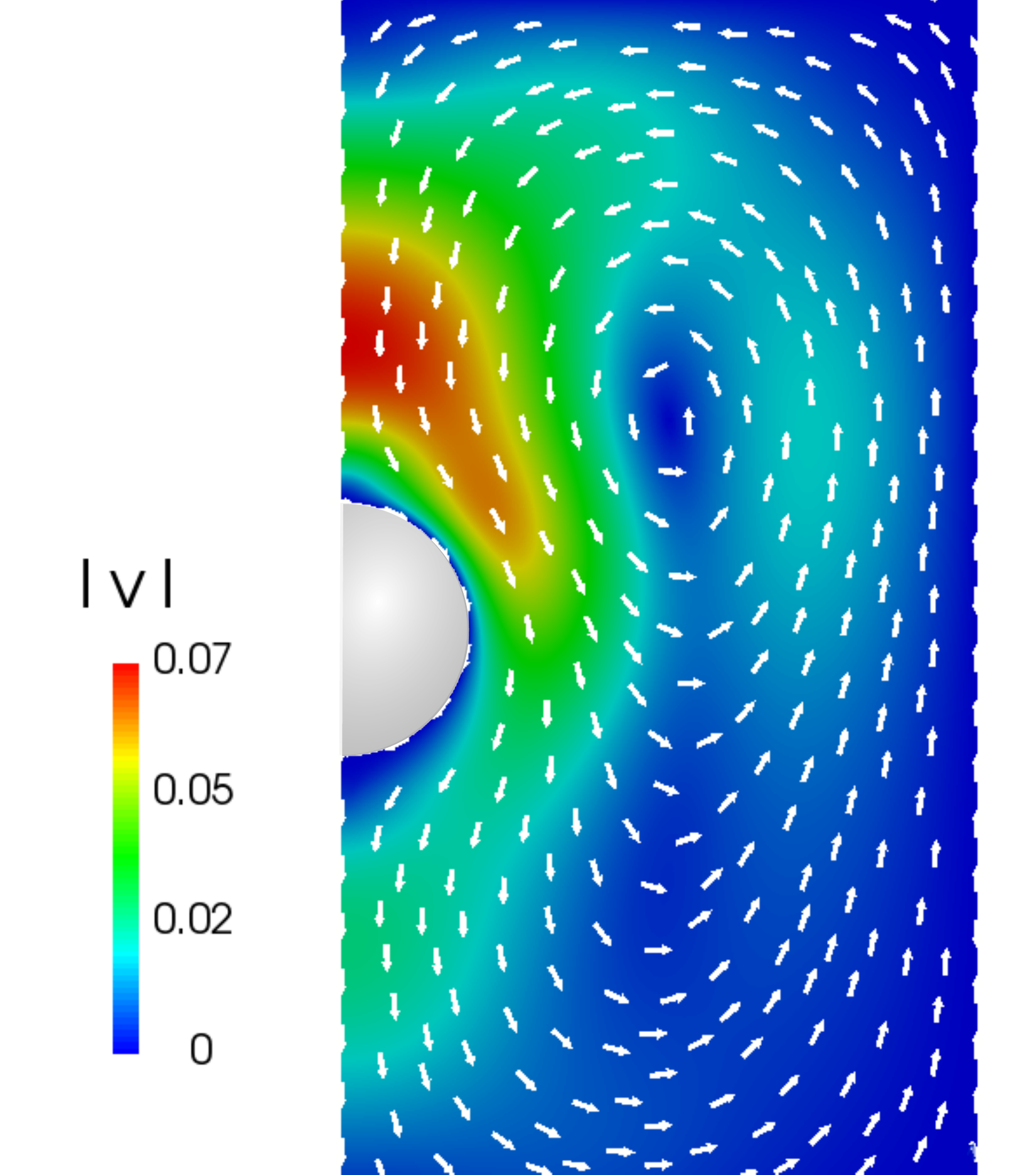}
\caption{Charge density and velocity of LCEO around the particle with a hedgehog generated by the electric field of inverted polarity.
Here $\lambda_{\varepsilon}=1$, $\lambda_{\sigma}=2$, $\tilde{\zeta}_1=0.3$, $\tilde{\zeta}_2=0$, $\tilde{\zeta}_4=1.3$, $\tilde{\zeta}_6=-0.15$.}\label{Reversed_field_plot}
\end{center}
\end{figure}

It is also worth noting that, if the applied electric field reverses, the charge distributions depicted in Fig.~\ref{Charge_plot} is inverted while the flow profiles shown in Fig.~\ref{Velocity_plot} remain unaltered (compare, for instance, Fig.~\ref{Reversed_field_plot} to Fig.~\ref{Charge_plot}b and Fig.~\ref{Velocity_plot}b).
 
We conclude that the differences between the flow profiles shown in Fig.~\ref{Velocity_plot} and the experimental observations in \cite{Lazo_2} are primarily due to different geometry of the experiment \cite{Lazo_2} where the electrolyte was confined to a planar cell of thickness comparable to the particle diameter. Furthermore, these differences stem from the fact that the particles considered in this study are much smaller than those in \cite{Lazo_2}. 

{\section{Conclusions}
In this paper we derived a mathematical model for electro-osmosis in nematic liquid crystals described in terms of the tensor order parameter. Following Onsager's variational approach to irreversible processes, we use the formalism that balances conservative and frictional forces obtained by varying the appropriately chosen free energy and dissipation functionals. In the current study these are given by their established expressions for nematic liquid crystals and colloidal suspensions. To illustrate the capabilities of the model, we consider a relatively simple example of electro-osmotic flow around an  immobilized spherical particle. The physically relevant micrometer-size of the particle is chosen so that (a) the elastic energy minimizing nematic configuration contains disclination loops that can only be described within a tensor order parameter theory and (b) the equations of the governing system decouple, simplifying the computational procedure. 

The numerical simulations for these particles demonstrate that both induced-charge- and liquid-crystal enabled electrokinetic effects are simultaneously present in the nematic electrolyte. The quadrupolar flow profiles around the particle encircled by an equatorial Saturn ring are symmetric with respect to the plane of the defect, while the particle accompanied by a hedgehog gives rise to the velocity fields $\mathbf{v}$ of dipolar symmetry. Unlike the LCEO in patterned nematics which vanishes when $\lambda_{\varepsilon}$ and $\lambda_{\sigma}$ are equal, here we observe nonzero velocity field $\mathbf{v}$ even in the case of $\lambda_{\varepsilon}=\lambda_{\sigma}$. This effect is in line with the model developed for ICEO flows around dielectric spheres and it should become more pronounced with the decreasing radius of the particle. When the applied electric field is reversed, the charge distribution within the system is inverted, while the flow profiles remain unaltered, confirming that the LCEO velocity is proportional to the square of the applied field.
 
We attribute the differences between the flow profiles obtained in this work and the experimental observations in \cite{Lazo_2} to the fact that the particle in the experiment was much larger and the geometry of the experiment itself was different. Here the particle was assumed to be suspended in space filled with the nematic electrolyte with the uniform director orientation away from the particle. On the other hand, in \cite{Lazo_2}, the electrolyte was confined to a planar cell of thickness comparable to the particle diameter.  

The proposed model can be also employed to study general electrokinetic phenomena in nematics, including the systems that contain macroscopic colloidal particles and complex network of topological defects.}

\begin{acknowledgments}
Support from the following National Science Foundation Grants is acknowledged by the authors: No. DMS-1434969 (D.~G. and O.~M.~T.), No. DMS-1435372 (C.~K., M.~C.~C., and J.~V.), No. DMS-1434185 (O.~L.), No. DMS-1434734 (N.~J.~W.), and No. DMS-1418991 (N.~J.~W.). The authors wish to thank Douglas Arnold for useful discussions regarding numerical simulations.
\end{acknowledgments}

\bibliography{e-osmosis}

\end{document}